\documentclass[fleqn,usenatbib]{mnras}

\usepackage[T1]{fontenc}
\usepackage{ae,aecompl}

\usepackage{graphicx}	
\usepackage{amsmath}	
\usepackage{amssymb}	
\usepackage{graphicx}	
\usepackage{amsmath}	
\usepackage{amssymb}	
\usepackage{breqn}
\usepackage{gensymb}

\newif\ifAMStwofonts
\newcommand{\be}{\begin{equation}}
\newcommand{\ee}{\end{equation}}
\newcommand{\bea}{\begin{eqnarray}}
\newcommand{\eea}{\end{eqnarray}}

\title[Evolving Galactic Dynamos \& Fits to NGC~4631]{Evolving Galactic Dynamos and Fits to the Reversing Rotation Measures in the Halo of NGC~4631}

\author[A. Woodfinden et al.]{Alex Woodfinden,$^{1}$\thanks{Email: a.woodfinden@queensu.ca}
R. N. Henriksen,$^{1}$
Judith Irwin,$^{1}$
and \newauthor  Silvia Carolina Mora-Partiarroyo$^{2}$\thanks{Email: silvia.carolina.mora@gmail.com}
\\
$^{1}$Dept. of Physics, Engineering Physics \& Astronomy, Queen's University, Kingston, Canada, K7L  3N6\\
$^{2}$Guest Researcher, Dept. of Physics, Engineering Physics \& Astronomy, Queen's University, Kingston, Canada, K7L  3N6
}

\date{Accepted XXX. Received YYY; in original form ZZZ}

\pubyear{2018}

\begin{document}
\label{firstpage}
\pagerange{\pageref{firstpage}--\pageref{lastpage}}
\maketitle

\begin{abstract}
Rotation measure (RM) synthesis maps of NGC~4631 show remarkable sign reversals with distance from the minor axis in the northern halo of the galaxy on kpc scales.  To explain this new phenomenon, we solve the dynamo equations under the assumption of scale invariance and search for rotating logarithmic spiral solutions.  Solutions for velocity fields representing accretion onto the disk, outflow from the disk, and rotation-only in the disk are found that produce RM with reversing signs viewed edge-on.  Model RM maps are created for a variety of input parameters using a Faraday screen and are scaled to the same amplitude as the observational maps.  Residual images are then made and compared to find a best fit.  Solutions for rotation-only, i.e. relative to a pattern uniform rotation,  did in general, not fit the observations of NGC~4631 well.  However, outflow models did provide a reasonable fit to the magnetic field.  The best results for the specific region that was modelled in the northern halo are found with accretion.  Since there is abundant evidence for both winds and accretion in NGC~4631, this modelling technique has the potential to distinguish between the dominant flows in galaxies.

\end{abstract}

\begin{keywords}
Dynamo -- Galaxies: Haloes -- Galaxies: Magnetic Fields -- Galaxies: Spiral
\end{keywords}



\section{Introduction}
\label{section:introduction}
Recent radio continuum observations of edge-on galaxies have revealed remarkable results.  Although large-scale regular magnetic field structures have been observed before in galaxy halos (eg. X-type fields, see \citet{Yelena2019}, \citet{Krause2006} and examples below), it is only recently that observational data have allowed us to probe the magnetic field component \textit{parallel to the line of sight} via rotation measures (RMs) in faint galactic halos.  RM synthesis \citep{Bren2005} has ensured that the data can be fully exploited to best advantage.  The physical quantity of interest is the Faraday depth which is the product of the line of sight component of the magnetic field, $B_\parallel$, (the `parallel' magnetic field), and the electron density, $n_e$.  The parallel field can be positive or negative depending on whether it points towards or away from the observer, respectively\footnote{See Sect.~2.3 of \citet{Yelena2019} for more details on RMs and how they are determined.}.  

          In Fig. \ref{fig:Carolina}, we reproduce Fig. 16 from from \citet{Carolina2018} (see also Fig. 6.10 from \citet{Mora2016}) showing a Faraday depth map, produced using RM synthesis, of the edge-on galaxy, NGC~4631, which has a strong, well-known halo.  In this figure, blue represents negative Faraday depths and red represents positive Faraday depths.  Consequently, the direction of the  magnetic field weighted and integrated  along the line of sight points away from the observer (blue) or towards the observer (red).  As can be seen, in the northern halo (on which a box has been drawn) there are regular {sign} reversals of the Faraday depth as one scans in the east-west direction.  These  {sign} reversals  are naturally explained by a regular halo magnetic field  that is alternating its azimuthal direction  on kpc scales in the galaxy. \textit{This is a new phenomenon, never before {\bf observed} in the halo of a galaxy.}

                    In the following, we refer to \textit{magnetic field reversals} when we refer to this observational phenomenon and this paper attempts to explain those reversals (see below).  Similar results have been seen in the \textit{disk} of the face-on galaxy, NGC~628, as shown in Figs.~18 and 26 of \citet{MBH2017} and also more recently in the \textit{disk} of the edge-on galaxy, NGC~4666 \citep{Yelena2019}.  For the latter galaxy, the field direction also flips across the major axis of the galaxy.  However, prior to the NGC~4631 result, no such phenomenon was seen in galactic halos.  Many of the 35 edge-on galaxies observed in the CHANG-ES survey \citep{Irwin2012} also show clear magnetic field reversals in the Faraday rotation maps and will be the subject of future work.  Thus reversing magnetic fields may be a common characteristic of galaxies, although not seen prior to the CHANG-ES survey. 



A variety of  both empirical and dynamo models for the structure of magnetic fields exists; examples  include: \citet{Sun2008}, \citet{Jaffe2010}, \citet{Jansson2012}, \citet{Ferriere2014}, and \citet{Terral2017}.  These models  recreate magnetic fields in galaxies using various observations of the Milky Way as well as external galaxies.  While these models have had some success, the fits use various inputs that may not necessarily be related to ISM parameters.  They are motivated primarily by observations, but are not derived from first principles.

\begin{figure}
\begin{center}

{\includegraphics[width=0.3\textwidth]{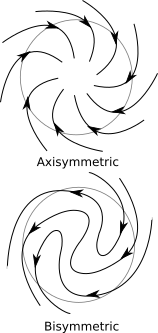}}
\caption{Examples of axisymmetric and bisymmetric field geometry.}  
\label{fig:symmetry}
\end{center}
\end{figure}

\begin{figure*}
{\includegraphics[width=0.9\textwidth]{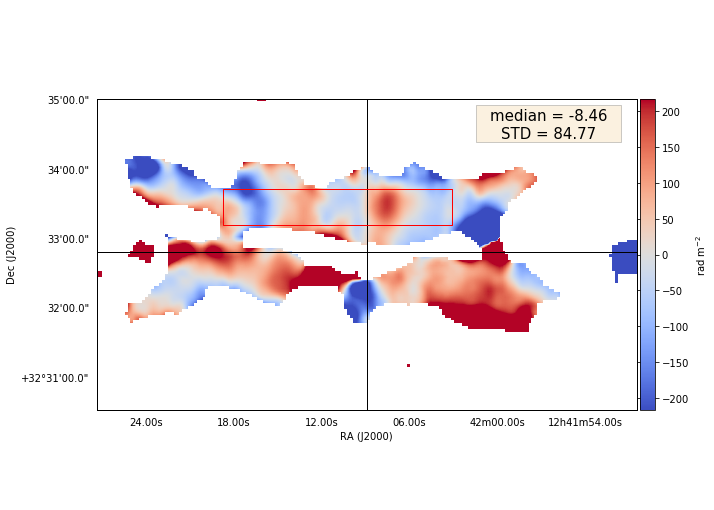}}
\caption{Distribution of Faraday depth obtained from C-band VLA (D array) data \citet{Carolina2018}. 
Faraday depth was clipped at 5$\sigma$ of polarized intensity. All data plotted have an angular resolution of 20.5$^{\prime\prime}$ FWHM.  The maximum error in this figure is about 80 rad/m$^2$ and decreases to about 20 rad/m$^2$ in areas of high polarization.  The red box displays regions showing magnetic reversals in the Northern Halo that is used in the analysis.  The median and standard deviation for this region is shown in the label on this figure and this region has a median error of 29.1 rad/m$^2$.  This figure was rotated by 5$\degree$ as per the position angle of NGC~4631 \citet{Mora2013}. The black lines on this figure show the major and minor axis of the galaxy, note however that there is some curvature to the major axis of this galaxy.  
}  
\label{fig:Carolina}
\end{figure*}

In recent work authors \citet{Terral2017} applied their empirical model to observations of the Milky Way to uncover the large scale magnetic field structure.  They found that the magnetic field in the galactic halo is more likely to be bisymmetric than axisymmetric (see Fig. \ref{fig:symmetry}).  This is because their bisymmetric model would show an X-shaped field if viewed externally and edge-on.  X type behaviour is well known from previous work for edge-on external galaxies \citep{Tullmann2000, Krause2006, Heesen2009, Braun2010, Soida2011, Haverkorn2012}.  It should be noted that the model used by the authors was limited  by the assumption that the magnetic field is non-helical when projected on cones.  X-shaped magnetic field structures is featured in a wide range of magnetic configurations showing spherical and quasi-spherical geometry \citep[e.g.][]{Brandenburg1992, Brandenburg1993}.

\citet{VanEck2015} used observations from 20 nearby galaxies to determine statistical properties of galactic magnetic fields and matched these with predictions of galactic dynamo theory.  Similar analysis was performed in \citet{Chamandy2016a} where pitch angles of observed galaxies are compared to $\alpha^2 \Omega$ dynamos and  reasonable agreement is found.  Papers such as \citet{Chamandy2016b} and \citet{Chamandy2014b} used various approximations such as saturisation of small time-scales to produce approximate solutions that are axisymmetric.  

The well studied dynamo theory \citep[e.g.][for a brief summary]{ KF2015} has made relevant predictions concerning X-type fields \citep[e.g.][]{Brandenburg1992} in halos and disks of galaxies and sign changes  in the halo as a function of height above the disk \citep[][and references therein]{Hen2017b}. However the theory is largely numerical and so is difficult to apply without intimate knowledge of the appropriate code.

In this paper, we replace many assumptions by the one assumption of scale invariance. The justification is that complex, self-interacting, dynamical systems frequently develop this symmetry \citep{Barenblatt96, Hen2015}. Moreover this assumption allows a relatively simple, semi-analytic, magnetic field description that is a solution of the classical dynamo equations. The ability to search through parameter space is illustrated by the multiple examples in Appendix~\ref{app:A}. In this sense we see it as a first step beyond the empirical models. Much of the detailed justification and comparison with earlier work is already included in \citet{Hen2017b} and \citet*{HWI2018}.

 Using the assumption of scale invariance  the  classical dynamo equations show again  that one can  produce  X-shaped magnetic fields, establish `parity' changes in a given halo quadrant,  and predict the field reversals in galaxy halos, as seen in NGC~4631.   The technique is similar to the study of axi-symmetric dynamos from \citet*{HWI2018}.  In the current paper we search for general azimuthal modes and so include both axially symmetric ($m=0$) and higher order modes. We find that a combination of axi-symmetric  and bi-symmetric modes ($m=1$, see Fig.~\ref{fig:symmetry} for the distinction) are required at minimum to  fit the various symmetries across quadrant boundaries. There are RM sign reversals in the same quadrant in the pure axially symmetric  mode \citep*[][Figs. 1 and 4]{HWI2018}, but  such reversals do not correspond to the multiple  regular RM reversals seen in Fig.~\ref{fig:Carolina}. This is  strong evidence for the bi-symmetric mode (or higher). While such magnetic field geometry has not so far been unambiguously detected in face-on galaxies \citep{Beck2015a, Beck2015b}, \citet{Fletcher2011} found that a bisymmetric spiral mode can fit observations of the face-on galaxy M51.  
 

In Sects.~\ref{section:spiral} and \ref{section:bi-symmetric}, we lay out the relevant theory within a self-similar framework.  Fields generated by classical dynamos are derived showing evolving  and rotating magnetic fields with different azimuthal modes. The fields tend to have spiral projections on cones about the minor axis as well as when projected onto the galactic plane. However the combined poloidal and toroidal structure of the field can be quite complex. Sample rotation measure (RM) screens for face-on galaxies are also presented.

In Sect.~\ref{section:fitting} we fit the RM screens produced by the evolving, scale invariant, magnetic fields to the Faraday rotation map of NGC~4631 seen in Fig.~\ref{fig:Carolina}.  We show the best fit results and the magnetic field that produces these fits.

Sect.~\ref{section:comparison} presents a comparison with previously published work and 
in Sect.~\ref{section:conclusions} we present our conclusions regarding the fits to NGC~4631.
          
In Appendix~\ref{app:A} we summarize the important physical results for face-on and edge-on cases.  These results will highlight the RM screens produced from a variety of velocity fields (e.g. inflow and outflow for a galaxy).  RM screens for both face-on and edge-on cases will be explored.              

\section{Scale Invariant, Evolving, Magnetic Dynamo Spiral fields}\label{section:spiral}

We refer to the classical mean-field dynamo equations  \citep{M1978} in the form for the  magnetic vector potential \citep{Hen2017b}
\be
\partial_t{\bf A}={\bf v}\wedge \nabla\wedge {\bf A}-\eta\nabla\wedge\nabla\wedge {\bf A}+\alpha_d\nabla\wedge {\bf A}.\label{eq:Afield}
\ee
A modern discussion of the limitations of this equation is sumarized in chapter 6 of \citet{ KF2015}. Scale invariance provides descriptions of the basic parameters $\alpha_d,~\eta,~ and ~{\bf v}$, but without the detailed physics. Scale invariant solutions are used in this work due to their simplicity, reproducibility, and ability to be easily tested against observational predictions.  The solutions  contain helicity that is present on all scales, \textit{which are coupled in time.}  It is important in the technical part of what follows to observe that the time derivative in this equation  is taken at a fixed spatial point. We do not therefore differentiate the unit vectors. 

In (\ref{eq:Afield}) ${\bf v}$ is the mean velocity, $\eta$ is the resistive diffusivity, $\alpha_d$ is the magnetic `helicity' resulting from  a helical sub-scale  magnetohydrodynamic velocity, and ${\bf A}$ is the magnetic vector potential.  The quantity $\alpha_d$ may be positive or negative \citep[e.g.][but we take it as positive in this work]{M1978}. Formally, $\eta$ is the Ohmic diffusivity $c^2/(4\pi\sigma)$ in terms of the electrical conductivity $\sigma$, but it can be interpreted also as a turbulent diffusivity of the form $\ell v_t$ given a turbulent velocity $v_t$ and spatial scale $\ell$. The sub-scales associated with the `helicity' and the `diffusivity' may not always be identical. 

Under the assumption of temporal scale invariance employed here, the  {\it amplitude} time dependence will simply be a power law or (in the limit of zero similarity class $a$ - see below) an exponential factor. Hence the spatial geometry of the magnetic field remains `self-similar' over the time evolution, and we can therefore study the geometry without requiring a fixed epoch. However our phase dependence includes a rotation in time (see the definition of the variables $\Phi$ and $\kappa$ below), which is an explicit description of `rotating magnetic spirals' in some background frame of reference.  Although the global geometry is self-similar,  any particular line of sight through the field may detect a different aspect of the spiral structure. The pattern angular velocity of the magnetic arms need not be the same as that of the stellar spiral arms. Indeed \citet{MBH2017} show that this is the case observationally.  This implies a time dependent phase difference between the two types of arms, which will only occasionally be zero. Should the magnetic spiral pattern speed be equal to that of the stellar arms, a constant phase shift is still possible. 

Our short hand  reference to `rotating magnetic spirals' is slightly misleading, as is our occasional reference to the field being `wound on cones'. In fact it is the projections of the field on cones symmetric about the galactic minor axis (including the galactic plane as a limiting such cone) that show spiral structure. The  three dimensional magnetic field is certainly not constrained to lie solely on cones (neither in fact is the vector potential),  as can be seen at lower left in Fig.~\ref{fig:wa1} below.

 The compatible time evolution of the quantities  $\alpha_d$, (and $\eta$ when  that is retained)  and  the mean flow velocity ${\bf v}$ is  also given by the scale invariance. This removes the necessity of arguing in detail about the physical origin of these quantities although their relative importance is an essential parameter. Ultimately these various  time dependences can be used to relate the current field amplitude to a `seed magnetic field', but we leave the restrictions on the value of this  seed field to another work.

The form of the scale invariance is found following  \citet{CH1991} and \citet{Hen2015}.  
We  introduce a time variable $T$ along the scale invariant direction according to 
\be
e^{\alpha T}=1+\tilde\alpha_d\alpha t,\label{eq:T}
\ee
where $\tilde \alpha_d$ is a numerical constant that appears in the scale invariant form for the helicity, $\alpha_d$, which form is to be given below.  The constant  numerical factor $\tilde\alpha_d$  in Eqn.~\ref{eq:T} is purely for subsequent notational convenience. The quantity $\alpha$ should not be confused with the helicity as it is an arbitrary reciprocal time-scale used in the  scaling. The cylindrical coordinates $\{r,\phi,z\}$ are transformed into scale invariant variables $\{R,\Phi,Z\}$\footnote{The exponential or power law {\textit {temporal}} scaling of these variables does not imply that the galactic variables (e.g. galactic radius) are also varying with time.  This scaling is only relevant to the {\textit {dynamo magnetic field}.}} according to \citet[e.g.][]{Hen2015}\footnote{We take spatial variables to be measured in terms of a fiducial unit such as the  radius of the galactic disk.} 
\be
r=Re^{\delta T},~~~~\Phi=\phi+(\epsilon/\delta +q \ln{r})\delta T,~~~~z=Ze^{\delta T},\label{eq:scaled variables}
\ee
where $\delta$ is another arbitrary reciprocal time-scale that appears in the spatial scaling, and $\epsilon/\delta$ is a number that fixes the rate of rotation of the magnetic field in time. We add $q$ to the arbitrary $\epsilon/\delta$ for subsequent algebraic convenience (see Eqn.~\ref{eq:kappa} below). 

It should once again be emphasized that the  quantities $\{R,\Phi,Z\}$  or some combination of these quantities, when used in the dynamo equations guarantee scale invariant solutions \citep[e.g.][]{Hen2015}. They are {\textit {not}} to be applied to the geometrical structure of the background galaxy. The  implications for the galaxy are through the forms required for the sub-scale helicity and diffusivity, as well as for velocities measured in some reference rotating frame.  These can be quite general in spatial form (see e.g. the comment after Eqn.~\ref{eq:params}, but they are reduced to functions of simple radius in this paper. 

Our theory does not give a value either for the rotational velocity of the magnetic field $\epsilon$ or for the magnetic spiral pitch angle,$1/q$ .  The latter seems to be similar to that of the stellar spiral arms while the magnetic pattern velocity may need considerations of outflow such as found in \citet{Moss2013} and \citet{Chamandy2014a}. In this latter connection if outflow above and below the disk arises from the active star formation part of the stellar arm (backside), then at less than the escape velocity it may lag the stellar arm to fall back somewhere behind the arm. This spiral arm  based `champagne flow' will create an amplified magnetic arm where it accretes. This will be at a phase shift relative to the stellar arm of roughly $\Omega_s~d/w$ where $w$ is the outflow velocity, $d$ is the radio scale height, and $\Omega_s$ is the pattern angular velocity of the stellar arm. If the pattern angular speed of the stellar arm is  much smaller than $w/d$, the magnetic arm should lag between multiple spiral arms.   

In our discussion $1/q>0$ appears as the tangent of the pitch angle of a  spiral mode that is lagging relative to the sense of increasing angle $\phi$\footnote{It should be noted that in \citet{Hen2017b}, $q$ had this role as the {\it normally} defined pitch angle with respect to the azimuth. In our examples $tan^{-1}(1/q)$ is typically $tan^{-1}(0.4)\approx 22^\circ$.}. 

We note from Eqn.~\ref{eq:T} that 
\be
e^{\delta T}=(1+\tilde\alpha_d\alpha t)^{1/a},\label{eq:delta}
\ee
where the `similarity class'  $a\equiv \alpha/\delta$ is a parameter of the model, which reflects the dimensions of a global constant. This quantity is discussed in some detail in \citet*{HWI2018}, but a simple example is afforded by a global constant $GM$ where $G$ is Newton's constant and $M$ is some fixed mass. This is the global constant for Keplerian orbits. 

Continuing with this special example, the space-time dimensions of $GM$ are $L^3/T^2$ and, after scaling length by $e^{\delta T}$ and time by $e^{\alpha T}$ \citep{CH1991}, $GM$ scales as $e^{(3\delta-2\alpha)T}$. To hold this invariant under the scaling we must set $\alpha/\delta\equiv a=3/2$, which is the `Keplerian  similarity class'.  Note that this `class', that is the ratio $3/2$ of the powers of spatial scaling to temporal scaling gives Kepler's third law, $L^3\propto T^2$ for any Keplerian motion. Similarly for a global constant with dimensions of velocity $a=1$, while a global constant with dimensions of specific angular momentum requires $a=2$. A constant angular velocity corresponds to $a=0$.  A tabular summary is provided in Table~\ref{table:a}.  

As is usual in this series of papers we write the magnetic field  for dimensional convenience as 
\be
{\bf b}=\frac{\bf B}{\sqrt{4\pi \rho}},\label{eq:b}
\ee
so that it has the dimensions of velocity. Here $\rho$ is a constant not associated with the dynamo and indeed might have the value $1/(4\pi)$ in cgs units, but it is completely arbitrary. It is in fact  absorbed into the  multiplicative constants that appear in our solutions.

In temporal scale invariance the  fields must have the following forms according to their dimensions 
\bea
{\bf A}&=&\bar{\bf A}(R,\Phi,Z)e^{(2-a)\delta T},\nonumber\\
{\bf b}&=& \bar{\bf b}(R,\Phi,Z)e^{(1-a)\delta T} \equiv e^{- \delta T} \Delta_{\bf X} \nabla {\bf A},\nonumber \\
{\bf v}&=& \bar{\bf v}(R,\Phi,Z)e^{(1-a)\delta T},\label{eq:tempfields}
 \eea   
where the barred quantities are the scale invariant fields, which are functions of the three scale invariant variables as defined in Eqns.~\ref{eq:scaled variables}. $\textbf{X}$ indicates that the cross product should be taken with respect to the scale invariant variables.  Eqns.~\ref{eq:Afield} can always be written solely in  terms of these scale invariant variables \citep{CH1991}, so that the temporal scaling symmetry eliminates only the $T$ dependence without additional assumptions. This is multi-variable scale invariance \citep{Hen2015,Barenblatt96}.

Considering Eqn.~\ref{eq:tempfields} and Eqn.~\ref{eq:delta} we see that the {\it amplitude} time dependence is generally a power law in powers of $(1+\tilde\alpha_d \alpha t)$, where the power is determined by the `class' parameter $a$.  Should $\alpha=0$ we find from Eqn.~\ref{eq:delta} that  $\delta T=\tilde\alpha_d \delta t$. The field can then  grow exponentially according to Eqns.~\ref{eq:tempfields}. The helicity, velocity field and indeed  the diffusivity will grow correspondingly. The time scale is controlled by the value of $1/(\tilde\alpha_d\delta)$, which may be long.
The helicity arising from the sub-scale $\alpha_d$, and the resistive diffusivity $\eta$, must be written  according to their respective dimensions as 
 \bea
 \alpha_d&=&\bar\alpha_d(R,\Phi,Z)e^{(1-a)\delta T},\nonumber\\
\eta&=& \bar \eta(R,\Phi,Z) e^{(2-a)T}.\label{eq:tempparams}
\eea 

At this stage a substitution of the forms Eqns.~\ref{eq:tempfields} into Eqns.~\ref{eq:Afield} yields three partial differential equations in the variables $\{R,\Phi,Z\}$.  However, we are seeking non-axially symmetric spiral symmetry in the magnetic fields to match the observations summarized in \citet{Beck2015a} and \citet{Kr2015}. Any combination of the  scale invariant quantities $\{R,\Phi,Z\}$ will  render the barred quantities in Eqns.~\ref{eq:tempfields} scale invariant, so we are free to seek a spiral symmetry by combining them. 

We choose a combination inspired by our previous modal analysis \citet{Hen2017b} and observations of `X-type' fields and magnetic spiral `arms'. We assume that the angular  dependence may be combined with $R$  in a  {\it rotating} logarithmic spiral form as  (recalling the definition of $\Phi$ from Eqn.~\ref{eq:scaled variables})
\be
\kappa\equiv \Phi+q\ln{R}\equiv \phi+q\ln(r)+\epsilon T.\label{eq:kappa}
\ee
Moreover we combine the $R$ and $Z$ dependence into  a dependence on the conical angle through 
\be
\zeta\equiv \frac{Z}{R}.\label{eq:zeta}
\ee
The linearity of Eqns.~\ref{eq:Afield} allows us to seek solutions in the complex form 
\be
  {\bf \bar A}(R,\Phi,Z)={\bf \tilde A}(\zeta)e^{im\kappa}.\label{eq:fieldform}
\ee
Note that the variable $\zeta$ is time independent. Hence the time dependence of the magnetic dynamo  appears only through the amplitude factors in Eqn.~\ref{eq:tempfields} and through the rotation of the modal pattern contained in the variable $\kappa$.

On substituting these assumed forms into Eqn.~\ref{eq:Afield} one finds that a solution is possible in terms of $\kappa$ and $\zeta$,  {\it provided that  the ancillary quantities satisfy }
\bea
\bar\alpha_d&=& \tilde\alpha_d\delta R,\nonumber\\
\bar\eta&=& \tilde\eta \delta R^2,\label{eq:params}\\
{\bf \bar v}&=&\tilde \alpha_d \delta R~\{u,v,w\}.\nonumber
\eea 
The quantities  denoted $\tilde{()}$ and the velocity components $\{u,v,w\}$ are dimensionless. They may at this stage be functions of the conical angle $\zeta$, but in the absence of definitive observations we keep these constant in this paper.

 Under these conditions the Eqns.~\ref{eq:Afield} become three linear equations for ${\bf A}(\zeta)$,
 \begin{dmath*}
(K+m^2\Delta)\tilde A_r-im\tilde A_z = (~1+~imq)v\tilde A_\phi-(1+\zeta v)\tilde A'_\phi-w\tilde b_\phi + \Delta\big(\tilde b'_\phi-im[1+imq)\tilde A_\phi-\zeta \tilde A'_\phi]\big)
 \end{dmath*}
 
 \begin{align}
 (K-&im~v+\Delta(1+m^2q^2))\tilde A_\phi \nonumber\\ 
 &= -u(1+imq)\tilde A_\phi+(\zeta u-w)\tilde A'_\phi+im(w\tilde A_z+u\tilde A_r)\nonumber\\
 &+ \tilde b_\phi + \Delta\{\zeta(1-2imq)\tilde A'_\phi+(1+\zeta^2)\tilde A''_\phi \nonumber\\
 &+ im[\zeta\tilde A'_r-\tilde A'_z+(1-imq)\tilde A_r]\}\nonumber
\end{align}
 
\begin{align}
im&\tilde A_r+(K+\Delta m^2(1+q^2))\tilde A_z \nonumber \\
&=  (1+imq)\tilde A_\phi+(v-\zeta)\tilde A'_\phi+u\tilde b_\phi\nonumber\\
&+\Delta\{\zeta \tilde b'_\phi-im[\tilde A'_\phi+q(\tilde A'_r+\zeta\tilde A'_z)]\}                                                            \label{eq:explicitA}
\end{align}

Where the prime indicates differentiation with respect to $\zeta$ and 
\bea
K&\equiv& (2-a)+im(\epsilon+v).\label{eq:K}\nonumber\\
\Delta&\equiv& \frac{\tilde\eta}{\tilde\alpha_d}.\label{eq:physparameters}
\eea
Here $\Delta$ is the inverse of the definition used in \citet{Hen2017b} in order to treat it as small when we wish to neglect diffusion. It might be a function of $\zeta$ at this stage. We anticipate a bit by writing the equations with $\tilde b_\phi$ included explicitly (we could of course write the equations entirely in terms of ${\bf \tilde b}$ but then the resulting field is not guaranteed to be solenoidal). This substitution is for brevity, but also because $\tilde b_\phi$ figures explicitly in our method of reducing the equations. We have set $\epsilon/\delta\rightarrow\epsilon$ so that the latter is now dimensionless. The angular velocity of the magnetic spiral pattern is $\epsilon\delta$.

The magnetic field that follows from the curl of the potential takes the form (omitting the power law amplitude factor given in Eqns.~\ref{eq:tempfields}
\be
\bar{\bf b}=\frac{\tilde{\bf b}}{R}e^{(im\kappa)},\label{eq:bbar}
\ee
where
\begin{dmath}
{\bf\tilde b} = \{ im\tilde A_z-\tilde A'_\phi,~~\tilde A'_r+\zeta \tilde A'_z-imq\tilde A_z,\\ ~~~~(1+imq)\tilde A_\phi-\zeta\tilde A'_\phi-im\tilde A_r\},\label{eq:btilde}
\end{dmath}
\begin{dmath*} \equiv \{\tilde b_r,\tilde b_\phi, \tilde b_z\}\nonumber 
\end{dmath*}
Eqns.~\ref{eq:btilde}, \ref{eq:bbar}, and the second of Eqns.~\ref{eq:tempfields} together {\it give the complete time dependent magnetic field}. In Eqn.~\ref{eq:btilde} $\tilde b_\phi$, as used in Eqns.~\ref{eq:explicitA}, is given explicitly in terms of the vector potential. 

Eqns.~\ref{eq:explicitA} are a complicated set of three linear ordinary equations with non constant coefficients. In general this is a numerical problem of at least fourth order. However the equations simplify to a second order equation when $\Delta=0$. This may be thought of as the zeroth order term in an expansion in $\Delta$, and so we proceed with this special case in this paper.  The resulting equations (Eqns.~\ref{eq:explicitA} with $\Delta=0$) reduce to the equations used in \citet*{HWI2018} for the axially symmetric temporal case when $m=0$.

An examination of Eqns.~\ref{eq:explicitA} with $\Delta=0$ indicates that one can rewrite Eqns.~\ref{eq:explicitA} as one second order equation for $\tilde A_\phi$. The algebra is however formidable. One effective procedure is to solve the second equation for $\tilde b_\phi$ in terms of $\tilde A_\phi$ and its derivatives. Then a substitution into the first and third equations yields two linear equations for $\tilde A_r$ and $\tilde A_z$ in terms of $\tilde A_\phi$ and its derivatives. These can be solved for $\tilde A_r$ and $\tilde A_z$, which are then to be substituted into the form of $\tilde b_\phi$ given in Eqn.~\ref{eq:btilde}. Finally this now independent expression (Eqns.~\ref{eq:explicitA}) does not know the form of $\tilde b_\phi$) for $\tilde b_\phi (\tilde A_\phi)$ is substituted into the second of Eqns.~\ref{eq:explicitA} to get a second order equation in $\tilde A_\phi$.
 
 The resulting equation is rather elaborate in general and we will only use  it in various special cases. We give instead the result  {\it before} the final substitution into the $\phi$ equation of Eqns.~\ref{eq:explicitA} as the two respective  equations for $\tilde b_\phi$
\begin{align}
&\tilde b_\phi (K^2-m^2(1+u^2+w^2)) \nonumber\\
&= (K-imv)\big[(K^2-m^2+(Ku-imw)(1+imq))\tilde A_\phi \nonumber\\
&+((w-u\zeta)K+im(u+w\zeta))\tilde A'_\phi\big],\label{eq:bphi1}
 \end{align}
\begin{align}
 &\tilde b_\phi\big(K^2 - m^2(1+qw)+imqKu\big)+\tilde b'_\phi\big((K-im\zeta)w\nonumber \\
 &-(K\zeta+im)u\big)= - (K-imv)\big((1+\zeta^2)\tilde A''_\phi\nonumber\\
 &-2imq\zeta\tilde A'_\phi+imq(1+imq)\tilde A_\phi\big).\label{eq:bphi2}
\end{align}

We emphasize that the second equation does not `know'  that the combination of potentials from Eqn.~\ref{eq:btilde} is in fact the azimuthal field. 
One must thus exercise caution in using these two equations. Rather than treating them as two equations for the quantities $\tilde A_\phi$ and $\tilde b_\phi$ , the correct procedure is to solve them simultaneously and substitute the first into the second in order to obtain a second order differential equation for $\tilde A_\phi$.  The resulting equation  is elaborate given a general velocity field as noted above, so that it is more convenient to make the substitution {\it after} a particular velocity field has been chosen.  

Subsequently the potentials $\tilde A_r$ and $\tilde A_\phi$  can be found from the first and third equations of Eqns.~\ref{eq:explicitA}. After eliminating $\tilde b_\phi$ and setting $\Delta=0$ these take  the forms 

\begin{dmath}
(K-imuw)\tilde A_r-im(1+w^2)\tilde A_z = [(1+imq)(v-uw)-w(K-imv)]\tilde A_\phi-[1+v\zeta+w(w-u\zeta)]\tilde A'_\phi,\label{eq:AR}
\end{dmath}
and 
\begin{dmath}
im(1+u^2)\tilde A_r+(K+imuw)\tilde A_z=[(1+imq)(1+u^2)+u(K-imv)]\tilde A_\phi +[v-\zeta+u(w-u\zeta)]\tilde A'_\phi.\label{eq:AZ}
\end{dmath}
Once again we leave the explicit linear solution for $\tilde A_r$ and $\tilde A_z$ for specific cases of the velocity field. Once these are found in terms of the solution for $\tilde A_\phi$  (Eqn.~\ref{eq:bphi2} after substituting Eqn.~\ref{eq:bphi1}),
all of the magnetic field components (including the azimuthal component in terms of $\tilde A_r$ and $\tilde A_z$) follow from the expressions in Eqns.~\ref{eq:btilde} and \ref{eq:bbar}.
In Sect.~\ref{sect:generic} we give a series of time dependent examples that are of interest in making qualitative comparisons with observations. One simplification  that is apparent from Eqns.~\ref{eq:AR} and \ref{eq:AZ} assumes  
the vertical velocity to vary on cones according to $w=u\zeta$.  This does not change Eqns.~\ref{eq:AR}, \ref{eq:AZ}, or the intermediate equation, Eqn.~\ref{eq:bphi1}, but the  equation, Eqn.~\ref{eq:bphi2}, for $\tilde A_\phi$ adds the term 
\be
(K-im\zeta)u,\label{eq:term}
\ee
to the bracket multiplying $\tilde b_\phi$.

\subsection{Boundary conditions}

The scale invariance of our solutions does not permit boundary conditions in $\zeta$, although the solutions behave fairly naturally there. However the galactic disk is essential to our study and generally it is not recognized by our solutions either. To obtain a solution valid for all $|z|$ we must impose a certain symmetry on the solution at the disk that is  taken to lie at $z=0$. Normally we impose a `dipolar' symmetry \citep[e.g.][]{KF2015} in which $B_z$ is held continuous across $z=0$ but $B_r$ and $B_\phi$ change sign after crossing $z=0$.

Formally, that is to embed  numerically the boundary condition into the solutions, Eqns.~\ref{eq:btilde} requires for the dipole symmetry that $A_z$ change sign across $z=0$ while $A_r$ and $A_z$ do not. In addition all derivatives of ${\bf A}$ should vanish at $z=0$. In practice we obtain the lower solution from the upper solution by reflecting the upper solution in the disk plane and changing the sign of the field. This requires a surface current at $z=0$ because of the tangential discontinuity.

An alternate symmetry is `quadrupolar' symmetry \citep[e.g.][]{KF2015}. The upper solution is simply reflected in the disk plane without a sign change. this changes the sign of $B_z$ but not of $B_r$ or $B_\phi$. The two sides of the disk are really independent under this symmetry. Formally Eqn.~\ref{eq:btilde} now requires $A_r$ and $A_\phi$ to change sign while $A_z$ does not, and all the derivatives of ${\bf A}$ to vanish at $z=0$, but we proceed with the reflected upper solution to obtain the lower solution.

With either of the imposed symmetries, the velocity field must change the sign of its helicity relative to the $z$ axis taken perpendicularly away from the pane on each side. This keeps both the tangential velocity  components and the vertical velocity component (thanks to the change in direction of the $z$ axis) continuous across $z=0$.

\section{\bf Generic Scale Invariant Dynamo Magnetic Field Modes}\label{section:bi-symmetric}
\label{sect:generic}
We look at some simple cases  in this section that illustrate generic properties. Specific fits to observational data  require more extensive parameter searches and multiple modes. These are discussed at length in Sect.~\ref{section:fitting} that contains the principal results of this paper. The axi-symmetric mode has been discussed in detail in \citet*{HWI2018}.

In \citet{Hen2017b} the notion of a uniformly rotating `pattern frame' as the rest frame of the dynamo magnetic field was introduced. The pattern frame may    also be the systemic frame of the galaxy, in which case the absolute field rotation would  be set essentially by the parameter $\epsilon\delta$. Generally we may think of this  pattern frame of reference as the pattern speed of the gravitational  spiral arms, and then $\epsilon\delta$ measures the  rotation of the magnetic arms relative to this reference frame. In the previous section we speculated that there would be a lagging phase shift relative to the stellar pattern. This is dependant on there being outflow, and so we use this as the generic case.

\subsection{Outflow or Accretion in the Pattern Reference Frame}

In this section we restrict ourselves to $a=1$ and $u=v=0$ in the pattern frame.  This allows us to study outflow from, or accretion onto, the galactic disk, which is an important observational question. We envisage application in this section to nearly edge-on galaxies, but we also display the existence of magnetic spirals in face-on disks and wound on cones in the halo.
 
 The combination of  Eqn.~\ref{eq:bphi1} with  \ref{eq:bphi2}  yields (the algebra can also be carried out directly from Eqns.~\ref{eq:explicitA} following the procedure outlined in general above) for $\tilde A_\phi$ 
\begin{dmath}
(1+w^2+\zeta^2)\tilde A''_\phi+2(Kw-imq\zeta)\tilde A'_\phi+[K^2-m^2(1+q^2)-im(w-q)]\tilde A_\phi=0,\label{eq:Aphiwa1}
\end{dmath}
where now 
\be
K=K(a=1;v=0)\equiv 1+im\epsilon.\label{eq:K10}
\ee
This equation is not invariant under a change in sign of $\zeta$ and $w$ as we would wish for the solution to apply above and below the galactic disk. We will  instead have to reflect the solution at $\zeta>0$ across the equatorial plane (with a sign change to keep the vertical field continuous) in order to create a symmetrical relation below the disk.  We find that both components of the tangential magnetic field must be  anti-symmetric across the disk \citep[see also][]{Hen2017b}.  The solution is given in terms of hypergeometric functions. 
We use the MAPLE\footnote{www.maplesoft.com} default cuts in the complex plane  fore these functions because these are continuous onto the cut from above. There are conditions for the convergence of the hypergeometric series however, With $\epsilon<0$ these reduce to $\zeta^2<3(1+w^2)$, which normally allows the halo to be covered adequately.

The equations for the remaining potentials may be found from Eqns.~\ref{eq:AR} and \ref{eq:AZ} in the explicit forms

\begin{figure*}
\begin{tabular}{cc} 
\rotatebox{0}{\scalebox{0.5} 
{\includegraphics{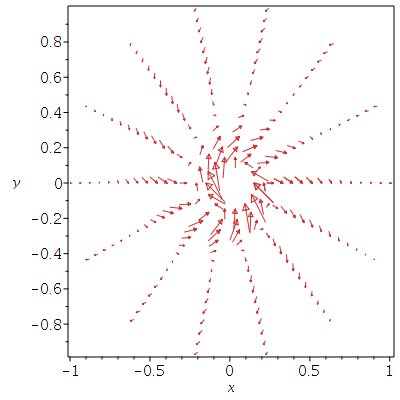}}}&
\rotatebox{0}{\scalebox{0.5} 
{\includegraphics{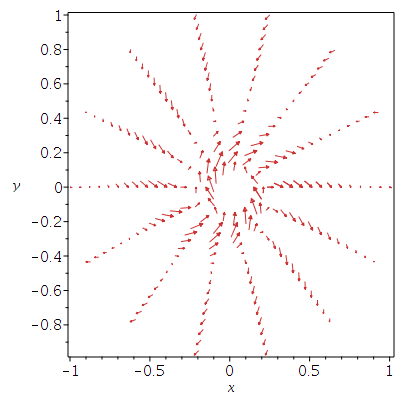}}}\\
{\rotatebox{0}{\scalebox{0.5} 
{\includegraphics{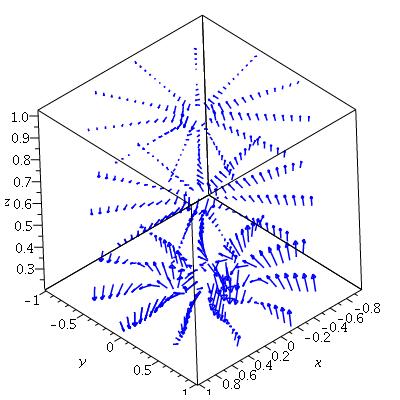}}}}&
\rotatebox{0}{\scalebox{0.65} 
{\includegraphics{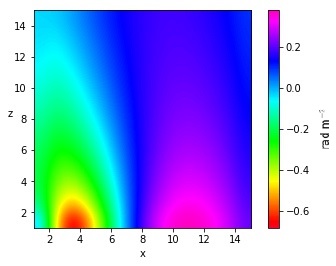}}}
\end{tabular}
\caption{These images are for the case with only $w\ne 0$ but positive and $a=1$.  At upper left the magnetic field vectors are shown on the conical surface $\zeta=0.5r$, while at upper right the field vectors are shown on a low vertical cut $z=0.15$. The radius of the galaxy is at $r=1$ in these units. In terms of  a parameter vector $\{ m,q,\epsilon,w,T,C1,C2 \}$, these plots have the vector $\{1,2.5,-1,2.0,1,1,0\}$. The radius runs over $0.15\le r \le 1$ in each case. Vectors are a fraction of the average at each point, with the maximum vector $0.5$ times the average value. The figure at lower left shows different slices in 3D for the same parameter vector as the previous images, but over the range $0.25\le r\le 1$. At lower right we show the rotation measure screen in the first quadrant, it should be noted that the scaling on these values is arbitrary and depends on a multiplicative constant. }  
\label{fig:wa1}
\end{figure*}

\begin{dmath}
[K^2-m^2(1+w^2)]\tilde A_r=[im(1+w^2)(1+imq)-K^2w]\tilde A_\phi-(1+w^2)(K+im\zeta)\tilde A'_\phi,\label{eq:Arwne0}
\end{dmath}
and
\begin{dmath}
[K^2-m^2(1+w^2)\tilde A_z=K[1+imq+imw]\tilde A_\phi-[K\zeta-im(1+w^2)]\tilde A'_\phi.\label{eq:Azwne0}
\end{dmath}
The dynamo magnetic field now follows from Eqn.~\ref{eq:btilde}. We show  some examples with simple parameter choices in Fig.~\ref{fig:wa1}.

\begin{figure*}
\begin{tabular}{cc} 
\rotatebox{0}{\scalebox{1.0} 
{\includegraphics{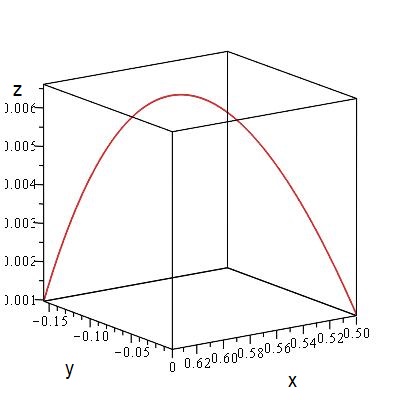}}}&
\rotatebox{0}{\scalebox{1.0} 
{\includegraphics{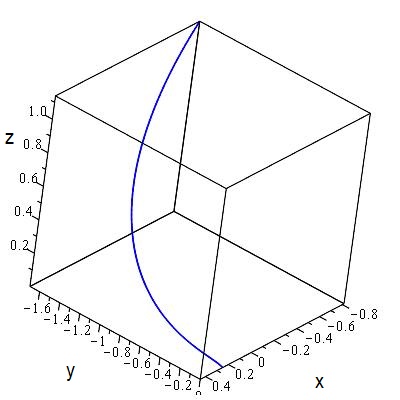}}}
\end{tabular}
\caption{ The closed magnetic field loop at left is for the same parameter set as in  the previous figure for $a=1$ and only $w\ne 0$. It begins at $\{r,\phi,z\}$ =$\{0.5,0,0.001\}$  and returns to the plane after looping in the spiral arm. The loop is very close to the plane with maximum at perhaps $60$ pc. The field line on the right is also for the same parameter set, but it begins closer to the centre at $\{r,\phi,z\}$ =$\{0.25,0,0.001\}$.  We see that this line descends (the field line is pointing downwards) from great heights while crossing over the centre of the galaxy.}    
\label{fig:fieldloopsW}
\end{figure*}

In Fig.~\ref{fig:wa1} we see a projected bi-symmetric spiral magnetic field. In principle the  projected spiral structure will continue to the centre of the galaxy, but with finite observational resolution the field might be seen there as a `magnetic bar'. The three dimensional field line structure is very markedly distributed in loops over the  projected arms. This may be detected  in the cube at lower left  and is confirmed in Fig.~\ref{fig:fieldloopsW}.  At small radius the field lines continue to great heights without looping as is  seen on the right in Fig.~\ref{fig:fieldloopsW}.  The cube at lower left  of Fig.~\ref{fig:wa1} also shows the field lines pointing towards the minor axis rather than away \citep{Kr2015}. \textit{In fact one normally finds the  diverging X-type  magnetic field only in the $m=0$ dynamo fields } \citep*[e.g.][]{HWI2018, Hen2017b}.
The rotation measure (RM) screen is shown in the first quadrant at lower right of Fig.~\ref{fig:wa1}, but the other quadrants may be generated by imposing anti-symmetry across the plane and either antisymmetry or symmetry across the minor axis depending on odd or even modes. We see that the RM changes sign mainly in radius, which suggests recourse to an $m=0$ axially symmetric component to achieve `parity inversion' with height. 
 
 We note that the magnitude of the outflow velocity is in terms of the turbulent velocity. This may be as high as $50$ km s$^{-1}$. So $w=2$ implies only a modest outflow. A value more like $w=5-10$ would be required to imitate the outflow velocities inferred elsewhere \citep{Heesen2018}. As may be expected, these tend to draw the magnetic field up into the halo and erase the parity change \citep{Henriksen2018}.
 
In Fig.~\ref{fig:fieldloopsW} we show  on the left a magnetic field line that loops very close to the plane inside the magnetic spiral. The parameters are the same as in Fig.~\ref{fig:wa1}. On the right we show a field line starting at smaller radii, but otherwise having the same set of parameters 
as on the left. The field line  extends to great heights and crosses over the centre of the galaxy.  It is important to note that these are not `Parker loops' 
arising from Parker instability, but are rather intrinsic to the magnetic dynamo. 

\begin{figure*}
\begin{tabular}{cc} 
\rotatebox{0}{\scalebox{0.5} 
{\includegraphics{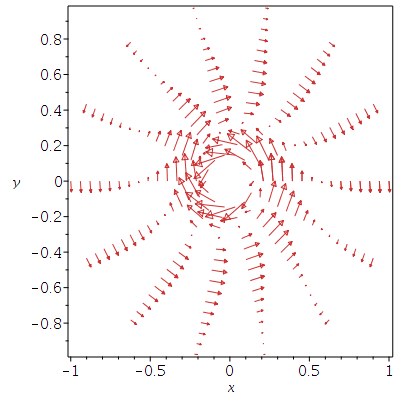}}}&
\rotatebox{0}{\scalebox{0.5} 
{\includegraphics{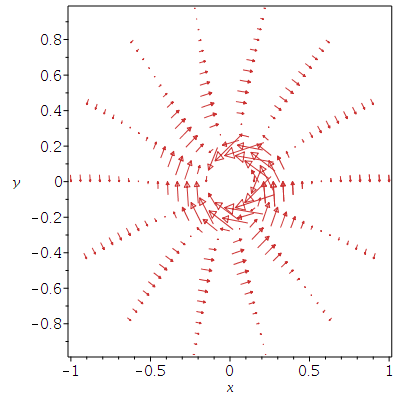}}}\\
{\rotatebox{0}{\scalebox{0.5} 
{\includegraphics{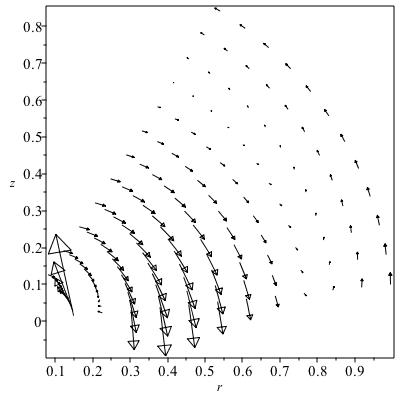}}}}&
\rotatebox{0}{\scalebox{0.65} 
{\includegraphics{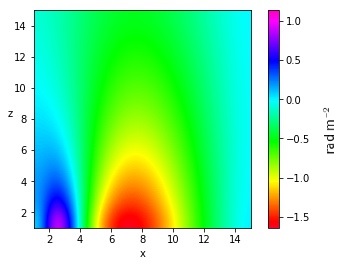}}}
\end{tabular}
\caption{The image at upper left of the figure shows a cut through the halo at $z=0.15$. The vertical velocity is $-2$ so that there is accretion onto the disk. The other parameters are the same as in Fig.~\ref{fig:wa1}, including the range of radius and $a=1$. At upper right we show the spiral structure on the cone $\zeta=0.25r$ over the same range in radius. Once again the only change is that the vertical flow is now inflow with $w=-2$.  At lower left we show a poloidal section at $\phi=\pi/4$. At lower right we show the RM screen for accretion  ($w=-2$) with the same parameters otherwise, it should be noted that the scaling on these values is arbitrary and depends on a multiplicative constant.  }    
\label{fig:accretionW}
\end{figure*}

The magnetic field is in fact stronger and the spirals are better defined under accretion ($w<0$) \citep{Hen2017}. This is demonstrated in Fig.~\ref{fig:accretionW}.

Fig.~\ref{fig:accretionW} shows a dramatic improvement of the projected magnetic spiral structure relative to the outflow results of Fig.~\ref{fig:wa1}, both at a constant cut in $z$ and projected onto the face of a cone.  At lower left we show a poloidal section at $\phi=\pi/4$ for the same accretion parameters. The field again loops above the disk, crossing over the centre of the galaxy (we have checked that the field at $\phi=5\pi/4$ has the opposite sign). The projected magnetic field is \textit{not} `X-shaped'. We have not corrected for  the  internal Faraday rotation of the locally produced emission in the presumed projections. 

The RM screen for the same accretion case is shown at the lower right of the Fig.~\ref{fig:accretionW}. Although the amplitudes vary considerably, most of the high halo is of uniform sign. the strong RM extends to greater heights than with the outflow.  Near the plane and near the minor axis  there is a strong sign change. Rapid variation in the magnetic field is also detectable in the poloidal section at lower left of the figure. A detailed Faraday depth model would require assuming the distribution of the relativistic electrons and ideally, performing RM synthesis (or the equivalent). We are only  calculating an RM screen, due solely to the magnetic field structure while assuming a constant electron density. Should both of these increase strongly with decreasing radius, our calculation mainly reflects conditions near the tangent point of the line of sight to a given circle in the disk. 

\begin{figure*}
\begin{tabular}{cc} 
\rotatebox{0}{\scalebox{0.75} 
{\includegraphics{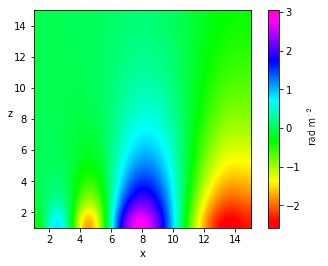}}}&
\rotatebox{0}{\scalebox{0.5} 
{\includegraphics{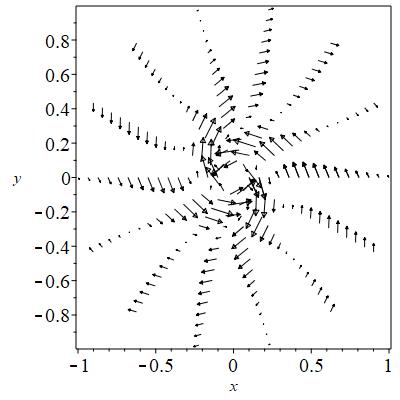}}}
\end{tabular}
\caption{ The figure shows the RM screen in the first quadrant for the parameter set $\{ m,q,\epsilon,w,T,C1,C2 \}$=$\{2,2.5,-1,-2,1,1,0\}$ in the left panel.  Scaling in arbitrary and depends on a multiplicative constant.  The sign change is now more frequent. The right panel is a cut at $z=0.15$ over the radial range $\{0.1,1\}$ for the same parameters, except that $q=1$.  }    
\label{fig:accretionm2RM}
\end{figure*}

In Fig.~\ref{fig:accretionm2RM} we show on the left the higher order mode $m=2$ for otherwise the same parameters as the accretion case in Fig.~\ref{fig:accretionW}. On the right we show the magnetic projected spiral structure for $m=2$ and $q=1$, a much larger pitch angle.

The RM  screen is  more structured because of the increased number of magnetic spirals in projection. The RM sign reversals continue from the disk into the halo although much of the activity is at small $\zeta$ (but moderate height). This type of oscillation in the RM was predicted  in \citet{Hen2017b} for modal solutions, and is confirmed here. The lack of resistivity in this analysis has not changed this behaviour very much, and so this behaviour may be generic to self-similar symmetry.

On the right hand panel of the figure we show a cut of the same example with accretion, but with a $45^\circ$ pitch angle.  This may be compared to the upper right panel in Fig.~\ref{fig:accretionW} with pitch angle $21.8^\circ$.  Similar behaviour is shown in the lower right panel of Fig.~1 in \citet{Hen2017b}, but again for  pitch angle $21.8^\circ$. Although we have made no attempt at a proper fit, these figures show a resemblance to the observations of NGC~4736 reported in Fig.~2 of \citet{CB2008}. The current example is for the class $a=1$ with infinite conductivity, while the example in \citet{Hen2017b} contains finite resistive diffusion and is for the similarity class $a=2$.  The velocity field, helicity and diffusion \citep[in][]{Hen2017b} all have global variations consistent with the specified $a$.  This particular galaxy is unique only in that it shows a two-armed mode extending well into the galactic centre independent of gravitational spirals. Many similar cases of magnetic spirals exist \citep{Beck2015a, WI2015}.

\begin{figure*}
\begin{tabular}{ccc} 
\rotatebox{0}{\scalebox{0.75} 
{\includegraphics{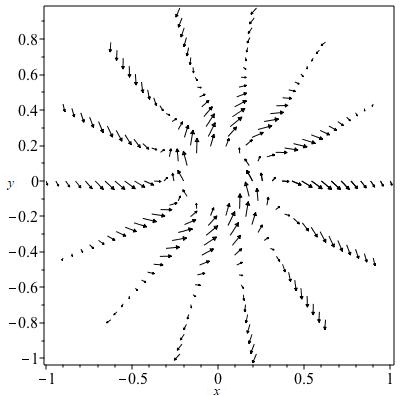}}}&
\rotatebox{0}{\scalebox{0.75} 
{\includegraphics{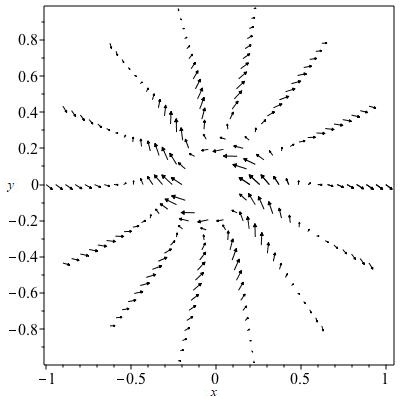}}}
{\rotatebox{0}{\scalebox{0.75} 
{\includegraphics{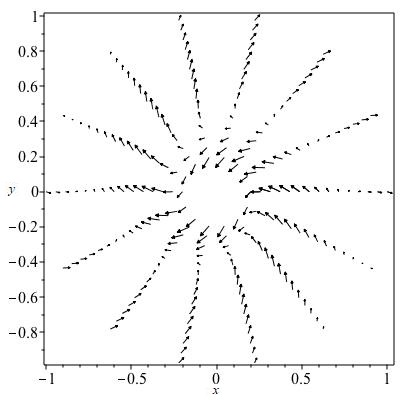}}}}&
\end{tabular}
\caption{The figure on the left is a cut through the solution of Fig.~\ref{fig:wa1} at $z=0.25$  but with $\{ m,q,\epsilon,w,T,C1,C2 \}=\{1,2.5,0,2,1,1,0\}$, so that  it has been rotated clockwise  through one radian relative to the figure at upper right in Fig.~\ref{fig:wa1} (which is also at a slightly lower cut $z=0.15$). The figure in the middle has been rotated counterclockwise through $45^\circ$  relative to that at upper left, while the right figure has been rotated counter clockwise through $90^\circ$. The line of sight is from the bottom of each figure.} 
\label{fig:rotatingcuts}
\end{figure*}

It is not obvious how the spiral arm pattern will be intersected by the line of sight (los). In our figures  we have taken it to lie at about  $-90 ^\circ$ to the $x$ axis. In Fig.~\ref{fig:rotatingcuts} we illustrate the changes that may be produced by this degree of freedom. We actually rotate the field pattern relative to the line of sight direction, which may be taken at the bottom of each figure.

Fig.~\ref{fig:rotatingcuts} shows the effect of rotating a spiral pattern relative to the los. This will appear strongly in the structure of the RM screen, which we do not include explicitly here for brevity. However the  qualitative differences  between the three cases in the integration of the parallel field along each los  starting from the bottom,  is evident by eye. Explicit examples are given in Appendix~\ref{app:A}.

\subsection{RM Screen for Face-on Galaxies}
\label{sect:faceon}

The previous section has demonstrated the existence of projected magnetic spirals in the disk and halo of a galaxy with an operating classical dynamo. These have been observed using the polarized emission from face-on and edge-on disks. However it is becoming common place to give the Faraday depth by RM synthesis  for nearly face-on galaxies \citep[e.g.][]{Beck2015b, MBH2017}. Thus in this section we give a preliminary RM screen analysis of essentially the same model used in the previous section. We continue to hold the electron density constant but if this quantity is determined observationally, a direct comparison with Faraday depth measurements will be possible.

We take a simple case where the axially symmetric stellar galaxy is inclined at a small angle $i$ to the line of sight (los), and the $x$ axis in the galaxy is taken perpendicular to the los pointing along the major axis to the west (north up). This simplification produces a glitch in our calculations at $\phi=\pi/2, 3\pi/2$ but the plotting routine is able to smooth out this effect. Just as in Fig.~\ref{fig:wa1} we take $\epsilon=-1$ so that the magnetic pattern is rotated counter-clockwise by one radian. This is of no real consequence here since we calculate the RM screen over $2\pi$ radians. 

We use cylindrical coordinates relative to the minor axis of the galaxy to describe the magnetic field. These are the set $\{r,\phi,z\}$ at the top surface of the disk/halo, which is taken to be a cylinder of height $h$ and radius equal to that of the disk (taken to be $1$). Along the line of sight ($d\ell$ starting from $\ell=0$ at the top) we must calculate the new  cylindrical coordinates $\{R(\ell), \Phi(\ell),Z(\ell)\}$ to obtain the los magnetic field. This field is (taken positive along the los towards an observer - written here for the third or fourth quadrant)
\bea
b_{los}&=&-b_r(R(\ell), \Phi(\ell),Z(\ell))\sin{(\Phi(\ell))}\sin{(i)}\nonumber \\ &-&b_\phi(R(\ell),\Phi(\ell),Z(\ell))\cos{(\Phi(\ell))}\sin{(i)}\nonumber\\
&+&b_z(R(\ell),\Phi(\ell),Z(\ell))\cos{(i)},\label{eq:blos} \label{eq:losfield}
\eea
where
\bea
R(\ell)&=&r[1+\frac{2\ell}{r}\sin({\phi})\sin{(i)}+\frac{\ell^2}{r^2}\sin{(i)}^2]^{1/2},\nonumber\\
\Phi(\ell)&=&\arctan{(\frac{r\sin{(\phi)}+\ell\sin{(i)}}{r\cos{(\phi)}})},\nonumber\\
Z(\ell)&=& h-\ell\cos{(\phi)}.\label{eq:ellcoords}
\eea
Our calculations are done at small enough radius and inclination that we do not worry about edge effects.

\begin{figure*}
\begin{tabular}{cc} 
\rotatebox{0}{\scalebox{0.5} 
{\includegraphics{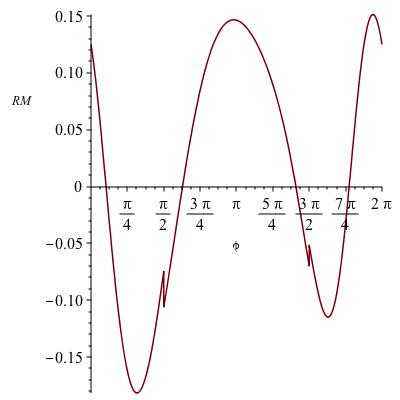}}}&
\rotatebox{0}{\scalebox{0.65} 
{\includegraphics{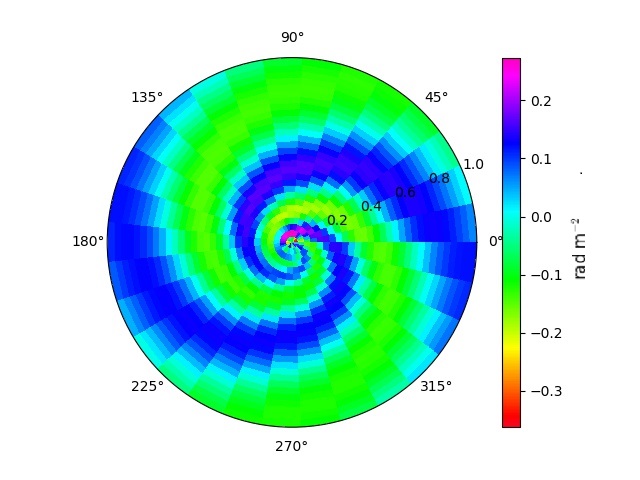}}}\\
{\rotatebox{0}{\scalebox{0.9} 
{\includegraphics{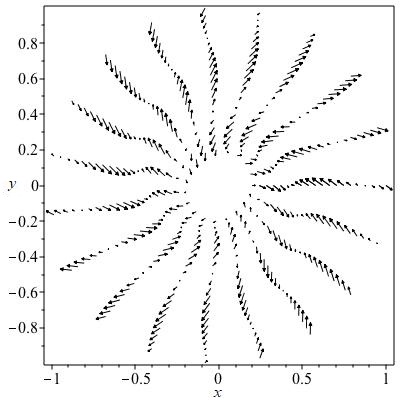}}}}&
\rotatebox{0}{\scalebox{1.4} 
{\includegraphics{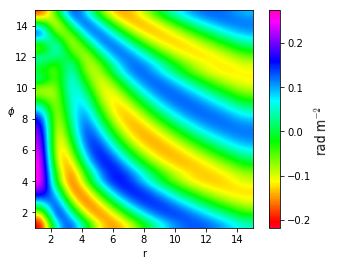}}}
\end{tabular}
\caption{ On the left of the figure we show the face-on RM for the parameter vector $\{r,i,h,m,q,\epsilon,w,T,C1,C2 \}=\{0.25,0.12,0.5,2,2.5,-1,2,1,1,0\}$.  That is a radial cut at $r=0.25$ over $0\le\phi\le 2\pi$. The figure on the right  shows in galaxy polar coordinates  the RM integrated over the face of the galaxy with the same parameters as on the left.  At lower left we show a cut along the los at $\ell=0.25$ for the same parameter set and at lower right we show the RM screen integrated over the face of the galaxy but in rectangular $\{r,\phi\}$ coordinates. The top of the figure fits smoothly on to the bottom of the figure and spiral structure is represented as inclined straight lines in the outer disk. The radius in the solution shown extends from 0 to 1 galactic radii and the angle extends from 0 to $2\pi$. Note that the colour bars at upper right and lower right are not the same and scaling is dependant on an arbitrary multiplicative constant.} 
\label{fig:faceonRM}
\end{figure*}

In Fig.~\ref{fig:faceonRM} we show on the left the integration of the magnetic field along the los  over $2\pi$ radians for a $m=2$ mode.  Because in these models the field tends to loop over the polarization arms, the RM maxima tend to be between and on the edges of the polarization arms.  The figure on the right shows the RM over the galactic plane in spherical polar coordinates. The spiral structure need not coincide with the polarization arms, although with the presence of the m=0 mode it may. By comparing the bottom two panels of the figure for the  pure $m=2$ mode, we infer that the central magnetic polarization arms are traced largely by the lines of nearly zero RM (light green colour in the figure).  Moreover it appears that the RM is negative on the inside of a polarization arm and positive on the outside of the arm. But this is highly model dependent and can be reversed by reversing the sign of multiplicative constants.  

In Appendix~\ref{app:A} we outline the observational expectations that result from systematically varying the parameters outlined in Sects.~\ref{section:spiral} \& \ref{section:bi-symmetric}.  We also summarize the physical interpretation of these parameters.

Similar face on magnetic behaviour {\it may} already have been detected in IC342 \citep{Beck2015b}. Other face on examples from our models are presented in Appendix~\ref{app:A}.

\section{Fit to NGC~4631}\label{section:fitting}

In this section we will fit RM screens generated from these dynamo models to the Faraday RM map of NGC~4631 from \citet{Carolina2018}.  This galaxy hosts one of the largest and brightest known galactic halos \citep{Wang2000, Wang2001} thought to be partly caused by gravitational interactions with neighbouring galaxies NGC~4565 and NGC~4627 \citep{Hummel1988, Mora2013}.  This merger is likely to have caused a starburst in the past leading to an outflow from this galaxy \citep{Irwin2011}.  This is a edge-on galaxy at an inclination of $89\degree \  \pm 1 \degree$ and an assumed distance of 7.6~Mpc \citep{Carolina2018}.  

Details of the observations and reductions used to create Fig.~\ref{fig:Carolina} can be found in greater detail in \citet{Carolina2018} and are briefly summarized below.  Observations were taken using the Karl. G. Jansky Very large Array (VLA) at C-band and L-band.  C-band data were selected as this is the only band at which one can expect to trace a large enough line of sight through the galaxy.  A map of the Faraday depth at a resolution of 20.5$^{\prime\prime}$ is created as shown in Fig.~\ref{fig:Carolina} of this paper.  The mid plane of the galaxy is completely depolarized and the median error in the region used for analysis is $29.1$ rad/m$^2$.  The Faraday rotation due to the galactic foreground is negligible in the direction of NGC~4631, \citet{Heald2009} found the galactic foreground to be $-4\ \pm\ 3$ rad/m$^2$ and \citet{Oppermann2012} found a value of $-0.3 \pm 2.7$ rad/m$^2$. Thus, the RM shown in Fig.~\ref{fig:Carolina} are intrinsic to NGC~4631.    

\citet{Heesen2018} looked at NGC~4631 as part of a sample of 12 galaxies.  They found a rotational velocity of $v_{rot} = 138$ km s$^{-1}$ \citep{Makarov2014} leading to an escape velocity of $v_{esc} = \sqrt{2} \times v_{rot} = 195$ km s$^{-1}$, where this is the escape velocity near the mid plane of the galaxy.  By fitting 1D cosmic ray transfer models they found an advection speed of $300^{+80}_{-50}$ km s$^{-1}$ in the northern halo and $200^{+50}_{-30}$ km s$^{-1}$ in the southern halo.  These values were typical of the other galaxies sampled.  The advection speed in the northern halo is clearly greater than the escape velocity and a net outflow from this galaxy is expected.  \textit{Due to different advection speeds in the northern and southern halos the dynamo solutions with the best fits are not expected to be the same above and below the disk.}  

The goal of fitting the dynamo solutions to the data is to explain the reversing sign of the RM seen in the northern halo of NGC~4631.  To do this a box is placed around the desired region as can be seen as the red box in Fig.~\ref{fig:Carolina}.  This box is chosen to encompass all of the reasonably regular reversals seen in the northern halo.  The uncertainty in the measurements is higher near the edges of the available data so the box is chosen to minimize this effect.  There is a strong reversal on the right of the halo seen as a dark blue patch in Fig.~\ref{fig:Carolina}, the strength of this reversal is several times higher than seen in other reversals and its shape is noticeably more rounded.  This reversal may not be due to the large scale field and may instead be another effect showing up in the rotation measure map such as a bubble.  As a precaution the box is chosen to avoid this region.

The dynamo solutions are re gridded to match the RM Synthesis map of NGC~4631.  The dynamo solutions are solved for up to one galactic radius on the major axis and one half galactic radius on the minor axis, the dynamo maps are resized to match NGC~4631 and properly centred to the galaxy.    As mentioned in Sects.~\ref{section:spiral} \& \ref{section:bi-symmetric} the dynamo solutions contain an arbitrary multiplicative constant that makes the strength of the magnetic field arbitrary.  To be able to fit these maps to the observation, the maps must be scaled to fit.  To do this the region inside the box selected in the northern halo of the galaxy is taken from both the observational and theoretical maps and the observation maps are divided by the theoretical maps.  The median of this new divided region is taken and used as a scaling factor.  The theoretical map is multiplied by this scaling factor.  This produces a new scaled dynamo map to match the scaling on NGC~4631.  

Once the new scaled dynamo maps have been created they are subtracted from the observed RM Synthesis maps of NGC~4631 to create residual maps that are then used to determine how well the dynamo field fit the observational results within the given region.  If the dynamo field matched the field of NGC~4631 the residual maps would be have a median of $0$ rad/m$^2$ and a standard deviation equal to the error in the image ($29.1$ rad/m$^2$).  These quantities as well as a goodness of fit test are used to compare how well different models fit the data.  The Akaike information criterion (AIC) is used as a goodness of fit test to estimate the relative quality of the models.  This is implemented using the procedure outlined in Sect.~4 of \citet{Erwin2015}, the lower the AIC value the better the model matches the data.  AIC estimates the quality of each model relative to other models given.  Thus, AIC provides a method for determining which model best represents the data. 

In order to determine the best fit dynamo and parameters a parameter search was done by calculating the dynamo solution for a large parameter space and then comparing each of these results to the observational map using the procedure outlined above.  For the outflow and accretion models the parameter $q$ was varied with the following values $q=\{0.0, 1.0, 2.5, 4.9, 11.5\}$ corresponding to pitch angles of $\{90 \degree, 45 \degree, 22 \degree, 13\degree, 5 \degree \}$.   The parameter $w$ was varied with the following values $w=\{2, 3, 4, 5, 6\}$ for the outflow case and the negative of these for the inflow case.  These values represent expected inflow and outflow velocities.   The rotation parameter $\epsilon$ was varied with the following values $\epsilon=\{-1.0, -0.5, 0.0, 0.5, 1.0\}$.   The parameter $m$ was varied with the following values $m=\{0,1,2 \}$ chosen to cover the first $3$ possible modes.   For pure rotation in the pattern frame the parameters $q$, $\epsilon$, $m$ were varied in the same manner as in the outflow/accretion cases.  The parameter $w$ was set to $0$ by requirement for the model.  The parameter $a$ was varied with the following values $a=\{0,1,2 \}$ (see Table~\ref{table:a}). 

From this parameter space the best results are shown in Table~\ref{table:nomixresults}.  These results were chosen because they are the only solutions which cause the standard deviation of the residual maps to be lower than the observational result.  As can be seen from this table more accretion and outflow solutions cause the residual maps to be closer to zero than rotation-only solutions.  These solutions, in general, also cause the standard deviation to be lower and have a lower AIC than the pure rotation cases. 

\begin{table*}
\centering
\begin{tabular}{llllll}
\hline
Model		& Case          & Parameter Vector                                                                    & Median & Standard Deviation & AIC \\ \hline \hline
1	& Outflow       & $\{m$, $q$, $\epsilon$, $u$, $v$, $w\}$ = $\{2, 2.5, 0.5, 0.0, 0.0, 3.0\}$          & 32.31  & 75.01    & 5909          \\
2	& Outflow       & $\{m$, $q$, $\epsilon$, $u$, $v$, $w\}$ = $\{2, 2.5, 0.5, 0.0, 0.0, 4.0\}$          & 7.20   & 71.39       &  4062     \\
3	& Outflow       & $\{m$, $q$, $\epsilon$, $u$, $v$, $w\}$ = $\{2, 2.5, 0.5, 0.0, 0.0, 5.0\}$          & -5.27  & 70.26    & 3707          \\
4	& Inflow        & $\{m$, $q$, $\epsilon$, $u$, $v$, $w\}$ = $\{1, 4.9, -1.0, 0.0, 0.0, -5.0\}$        & 11.51  & 82.12     & 4105         \\
5	& Inflow        & $\{m$, $q$, $\epsilon$, $u$, $v$, $w\}$ = $\{2, 2.5, 0.5, 0.0, 0.0, -2.0\}$         & 18.80  & 75.48     & 3308        \\
6	& Inflow        & $\{m$, $q$, $\epsilon$, $u$, $v$, $w\}$ = $\{2, 2.5, -1.0, 0.0, 0.0, -3.0\}$        & -28.47 & 72.30    & 3465          \\
7	& Inflow        & $\{m$, $q$, $\epsilon$, $u$, $v$, $w\}$ = $\{2, 2.5, 0.5, 0.0, 0.0, -3.0\}$         & 14.32  & 69.15     & 2862         \\
8	& Inflow        & $\{m$, $q$, $\epsilon$, $u$, $v$, $w\}$ = $\{2, 2.5, -1.0, 0.0, 0.0, -4.0\}$        & -25.12 & 65.09    & 2785          \\
9	& Inflow        & $\{m$, $q$, $\epsilon$, $u$, $v$, $w\}$ = $\{2, 2.5, 0.5, 0.0, 0.0, -4.0\}$         & 11.17  & 68.00    &   2818        \\
10	& Inflow        & $\{m$, $q$, $\epsilon$, $u$, $v$, $w\}$ = $\{2, 2.5, -1.0, 0.0, 0.0, -5.0\}$        & -19.65 & 61.08    & 2445          \\
11	& Inflow        & $\{m$, $q$, $\epsilon$, $u$, $v$, $w\}$ = $\{2, 2.5, 0.5, 0.0, 0.0, -5.0\}$         & 12.56  & 67.94     & 2869         \\
12	& Rotation-Only & $\{a$, $m$, $q$, $\epsilon$, $u$, $v$, $w\}$ = $\{0, 1, 4.9, -1.0, 0.0, 1.0, 0.0\}$ & -34.43 & 81.51 & 5191              \\
13	& Rotation-Only & $\{a$, $m$, $q$, $\epsilon$, $u$, $v$, $w\}$ = $\{0, 2, 1.0, -1.0, 0.0, 1.0, 0.0\}$ & 88.38  & 75.39 & 10262             \\
14	& Rotation-Only & $\{a$, $m$, $q$, $\epsilon$, $u$, $v$, $w\}$ = $\{1, 2, 2.5, 1.0, 0.0, 1.0, 0.0\}$  & 74.63  & 84.38  & 9272            \\ \hline
\end{tabular}%
\caption{Results of solutions where the standard deviation of the residual maps was less than the standard deviation of the NGC~4631 RM map in the specified box from Fig.~\ref{fig:Carolina}, indicating a good fit.  No mixing of different modes was allowed in this table and the mode number is specified in the parameter vector.  Left most column indicates the type of solution (inflow-only, outflow-only, or rotation-only).  The parameter vector for each solution is shown in the column second from the left.  The next right two columns indicate the median and standard deviation for the desired box in the solutions (see Fig.~\ref{fig:Carolina}).  The rightmost column shows the AIC indicating the goodness of fit for the models.}
\label{table:nomixresults}
\end{table*}

We can also combine different modes from the same dynamo models (with the same parameter set) together to create a new map to fit the observational results.  These maps do not have to be the same magnitude and the amplitude of each mode can be varied to account for certain modes being dominant.  The spiral pitch angle is also an additional degree of freedom that can be varied for the different modes however for this analysis it is assumed to be consistent between them.  In order to test these maps we take the parameter sets from Table~\ref{table:nomixresults} and allow the $m=0,1,2$ modes to mix.  To do this we create another large parameter space where the amplitudes of each of these modes is varied from the following values: $\{ -2.0$, $-1.5$, $-1.$, $-0.5$, $-0.1$, $-0.001$, $0.$, $0.001$, $0.1$, $0.5$, $1.$, $1.5$, $ 2.0 \}$.  The new maps created from these modes are then compared to the observational maps with the same procedure as above to determine the best fits. 

This is done and best results are shown in Table~\ref{table:mixresults}.  We show the parameter vector, median, and standard deviation within the reversal region in the northern halo, and the combined amplitudes that provide the lowest AIC value.  As can be seen, the outflow/inflow solutions again perform considerably better than the rotation-only results.  

\begin{table*}
\centering
\resizebox{\textwidth}{!}{%
\begin{tabular}{lllllll}
\hline
Model		& Case          & Parameter Vector                                                                    & Median & Standard Deviation & AIC & Mode Amplitudes\\& & & & $m=(0,1,2)$ \\ \hline \hline
1m	& Outflow       & $\{m$, $q$, $\epsilon$, $u$, $v$, $w\}$ = $\{m, 2.5, 0.5, 0.0, 0.0, 3.0\}$          & -17.56  & 76.86   & 4331           & (-0.1,0,0.5)            \\
2m	& Outflow       & $\{m$, $q$, $\epsilon$, $u$, $v$, $w\}$ = $\{m, 2.5, 0.5, 0.0, 0.0, 4.0\}$          & -2.3  & 69.21 & 3771              & (-0.1,-0.1,2.0)               \\
3m	& Outflow       & $\{m$, $q$, $\epsilon$, $u$, $v$, $w\}$ = $\{m, 2.5, 0.5, 0.0, 0.0, 5.0\}$          & -23.41 & 70.87 &3535             & (-0.5,-0.5,2.0)              \\
4m	& Inflow        & $\{m$, $q$, $\epsilon$, $u$, $v$, $w\}$ = $\{m, 4.9, -1.0, 0.0, 0.0, -5.0\}$        & -6.84  & 53.7 & 2082             & (-0.1,1.5,1.5)               \\
5m	& Inflow        & $\{m$, $q$, $\epsilon$, $u$, $v$, $w\}$ = $\{m, 2.5, 0.5, 0.0, 0.0, -2.0\}$         & 16.17  & 74.74  & 3262            & (0,0.001,0.1)               \\
6m	& Inflow        & $\{m$, $q$, $\epsilon$, $u$, $v$, $w\}$ = $\{m, 2.5, -1.0, 0.0, 0.0, -3.0\}$        & -6.11 & 74.45 & 3253             & (0.1,0,1.0)               \\
7m	& Inflow        & $\{m$, $q$, $\epsilon$, $u$, $v$, $w\}$ = $\{m, 2.5, 0.5, 0.0, 0.0, -3.0\}$         & 14.47  & 69.13 & 2859             & (0,0.001,0.5)              \\
8m	& Inflow        & $\{m$, $q$, $\epsilon$, $u$, $v$, $w\}$ = $\{m, 2.5, -1.0, 0.0, 0.0, -4.0\}$        & -14.83 & 66.04 & 2623              & (0.1,0,2.0)             \\
9m	& Inflow        & $\{m$, $q$, $\epsilon$, $u$, $v$, $w\}$ = $\{m, 2.5, 0.5, 0.0, 0.0, -4.0\}$         & 17.01  & 67.74 & 2800             & (-0.001,0.1,2.0)             \\
10m	& Inflow        & $\{m$, $q$, $\epsilon$, $u$, $v$, $w\}$ = $\{m, 2.5, -1.0, 0.0, 0.0, -5.0\}$        & -5.94 & 62.77 & 2346             & (0.1,0,1.5)             \\
11m	& Inflow        & $\{m$, $q$, $\epsilon$, $u$, $v$, $w\}$ = $\{m, 2.5, 0.5, 0.0, 0.0, -5.0\}$         & 16.06  & 67.22 & 2803              & (-0.001,0.1,1.5)              \\
12m	& Rotation-Only & $\{a$, $m$, $q$, $\epsilon$, $u$, $v$, $w\}$ = $\{0, m, 4.9, -1.0, 0.0, 1.0, 0.0\}$ & -33.53      & 81.17 & 4994                  & (-2.0,0.1,0)                            \\
13m	& Rotation-Only & $\{a$, $m$, $q$, $\epsilon$, $u$, $v$, $w\}$ = $\{0, m, 1.0, -1.0, 0.0, 1.0, 0.0\}$ & -1.53      & 101.19 & 7769                  & (-2.0,0.1,-0.1)                            \\
14m	& Rotation-Only & $\{a$, $m$, $q$, $\epsilon$, $u$, $v$, $w\}$ = $\{1, m, 2.5, 1.0, 0.0, 1.0, 0.0\}$  & -26.46      & 80.36 & 5232                  & (-0.5,0.001,0.001)                            \\ \hline
\end{tabular}%
}
\caption{Results of solutions where the standard deviation of the residual maps was less than the standard deviation of NGC~4631 (from Fig.~\ref{fig:Carolina}) in the specified box for the no mixing case.  Mixing of different modes was allowed in this table.  Left most column indicates the type of solution (inflow-only, outflow-only, or rotation-only).  The parameter vector for each solution is shown in the column second from the left.  Columns 3 \& 4 indicate the median and standard deviation for the desired box in the solutions (see Fig.~\ref{fig:Carolina}).  The rightmost column is the amplitudes fo the $m=0,1,2$ modes respectively.  Note these combinations are then rescaled to NGC~4631 before creating the residual map.}
\label{table:mixresults}
\end{table*}

\begin{figure*}
\begin{tabular}{cc} 
\rotatebox{0}{\scalebox{0.65} 
{\includegraphics{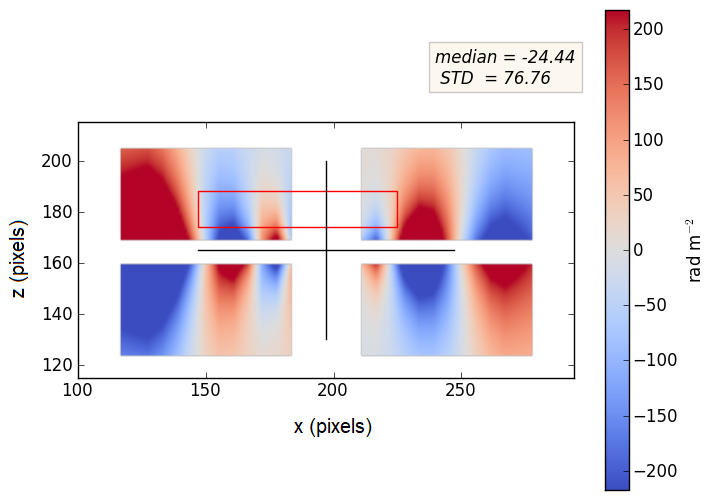}}}&
\rotatebox{0}{\scalebox{0.65} 
{\includegraphics{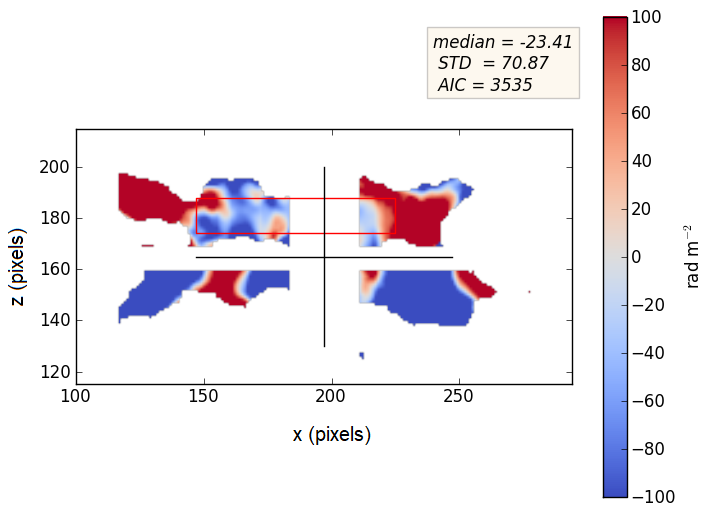}}}\\
\end{tabular}
\caption{We present an outflow-only solution that best matches the observed map of NGC~4631.  The scaled solution is shown on the left and the corresponding residual map is shown on the right corresponding to model $3$ in Table~\ref{table:nomixresults}.  Red box displays regions showing magnetic reversals in the Northern Halo of NGC~4631 that is used in the analysis.  The median and standard deviation for this region is shown in the label on this figure.  This solution is obtained with the parameter vector $\{m$, $q$, $\epsilon$, $u$, $v$, $w\}$ = $\{m, 2.5, 0.5, 0.0, 0.0, 4.0\}$ where we combined 3 modes ($m=0,1,2$) with scaling factors $(-0.5, -0.5, 2.0)$.  Once combined the solutions are rescaled to match the observed map in the method described in Sect.~\ref{section:fitting}.  Note the change of scale between the two figures.} 
\label{fig:bestoutflow}
\end{figure*}

The outflow solutions were improved through the combinations of different modes.  The best outflow model without combining modes was model $3$ with the parameter vector  $\{m$, $q$, $\epsilon$, $u$, $v$, $w\}$ = $\{2, 2.5, 0.5, 0.0, 0.0, 5.0\}$ which is a solution with moderate outflow wind speeds relative to other solutions.  The solution with the best fit that combines several modes is model $3m$ with parameter vector $\{m$, $q$, $\epsilon$, $u$, $v$, $w\}$ = $\{m, 2.5, 0.5, 0.0, 0.0, 5.0\}$ where we combined 3 modes ($m=0,1,2$) with scaling factors $(-0.5, -0.5, 2.0)$ respectively.  This fit is shown in Fig.~\ref{fig:bestoutflow}.  Note that once combined the solutions are rescaled to match the observed map in the method described above.  This solution has a moderate outflow velocity and the magnetic spirals have a pitch angle of $22 \degree $.  A pitch angle of $22 \degree$ is typical and velocities are in units of the subscale turbulent velocity.  A turbulent velocity value must be adopted to convert to physical units and a value of 50 km s$^{-1}$ leads to an outflow velocity of $\sim 250$ km s$^{-1}$ for $w=5$.  This compares favourably to the measured outflow velocity for NGC~4631 from \citet{Heesen2018} and is within the error range of their value.

\begin{figure*}
\begin{tabular}{cc} 
\rotatebox{0}{\scalebox{0.65} 
{\includegraphics{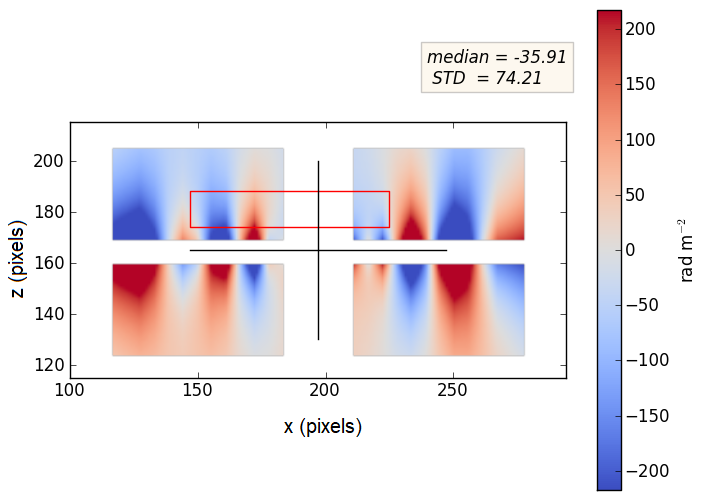}}}&
\rotatebox{0}{\scalebox{0.65} 
{\includegraphics{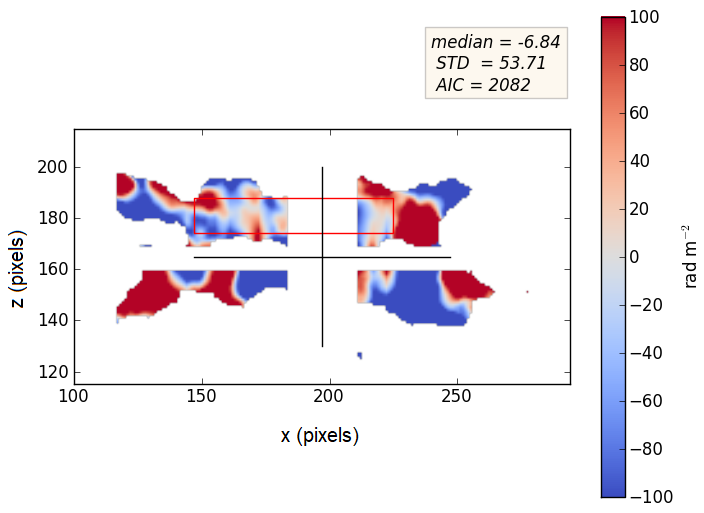}}}\\
\end{tabular}
\caption{We present an inflow-only solution that best matches the observed map of NGC~4631.  The scaled solution is shown on the left and the corresponding residual map is shown on the right corresponding to model $4m$ in Table~\ref{table:mixresults}.  The red box displays the region showing magnetic reversals in the Northern Halo of NGC~4631 that is used in the analysis.  The median and standard deviation for this region is shown in the label on this figure.  The solution is obtained with the parameter vector $\{m$, $q$, $\epsilon$, $u$, $v$, $w\}$ = $\{m, 4.9, -1.0, 0.0, 0.0, -5.0\}$ where we combined 3 modes ($m=0,1,2$) with scaling factors $(-0.1, 1.5, 1.5)$.  Once combined, the solutions are rescaled to match the observed map in the method described in Sect.~\ref{section:fitting}.  Note the change of scale between the two figures.}
\label{fig:accretioncombined}
\end{figure*}

\textit{The accretion solutions provide the best fit to the maps of NGC~4631 and an improvement to these fits is also seen when different modes are combined.}  As can be seen, the lowest standard deviation and AIC is provided from model $4m$ with parameter vector $\{m$, $q$, $\epsilon$, $u$, $v$, $w\}$ = $\{m, 4.9, -1.0, 0.0, 0.0, -5.0\}$ and has scaling factors for the $m=0,1,2$ modes of $(-0.1, 1.5, 1.5)$.  This solution is seen in Fig.~\ref{fig:accretioncombined}.  This is a solution that has a moderate to strong inflow velocity and magnetic spirals with a pitch angle of $11.5 \degree$.  The strongest modes are $m=1,2$ however the $m=0$ mode is present and non-negligible in the fit. 

These results show that the magnetic field of NGC~4631 can be well fit by scale invariant dynamo solutions with either accretion or outflow onto/from the galaxy.  This model in its present form imposes various constraints such as assuming accretion or outflow is proportional to the radius throughout the galaxy, the electron density is constant through the galaxy, and only large scale magnetic fields are seen in the observations, etc.  Despite these scale invariant requirements the magnetic field of NGC~4631 using the RM maps of the galaxy was well described by RM maps of dynamo models.  Dynamo solutions for rotation-only cases did, in general, not fit the observations of NGC~4631.

The fact that inflow and outflow models are quite similar makes it difficult to distinguish between the two (see Appendix~\ref{section:outflowaccretion}).  Nevertheless, there is a clear although marginal preference for our data to be better fit by infall models.  This result was unexpected since many authors have argued for winds from NGC~4631 as well as other galaxies  \citep[][and others]{Heesen2018, Hummel1990, Mora2013, Tullmann2006}.

The difference may be due to the restricted range over which our fits were carried out.  However, we note that the environment of NGC~4631 shows considerable complexity because of the well-known interaction with the galaxies NGC~4656 and NGC~4627.  Numerous HI spurs and tidal features are seen connecting these systems and there is also strong evidence for infalling gas \citep[for example, see][]{Combes1978, Stephens1990, Rand1994, 
  Richter2018}.

Our models therefore have the potential to provide an important discriminator between such scenarios especially as data improve and more such systems are observed.

\section{Comparison with Previous Models}
\label{section:comparison}

X-shaped fields are seen in many edge-on galaxies (see Sect.~\ref{section:introduction})
  and are predicted here for the $m=0$ mode, as well as in much earlier work \citep{Brandenburg1992, Brandenburg1993}. The latter two papers cited contain many of the same effects that we have found, although in axial symmetry.

 In \citep{Brandenburg1992} the dynamo equations are integrated numerically in space and time using rather detailed  assumptions regarding wind and rotational velocities, alpha effect and diffusivity. Moreover they introduce dynamo action in the halo much as do we. A significant result compared to our own findings is the complex variation with time  and angle of the RM, when projected onto the galactic plane as in our section \ref{sect:faceon}.  This is shown in their Fig.~5; the structure varies in time much as would our fields due to pattern rotation.  These authors also suggest complex parity structure in the halo, but they do not show the RM predicted for edge-on galaxies. In \citep{Brandenburg1993} the same type of integration is used to produce X-type fields in the halo in axial symmetry (their Figs. 8b, 8c).  It should be noted that we agree that the $m=0$ mode is required to produce the  X-type fields. 

The assumption of scale invariance that we use has the following advantages compared to the earlier insightful work. It offers a coherent assumption for the alpha effect in the halo, for diffusivity, and for rotational and wind velocity, which are not grossly unphysical. Because this  assumption renders the  solutions semi-analytic, they can be used relatively straightforwardly to fit observations as we have shown. Moreover scale invariance is a commonly occurring symmetry in complex systems and likely to be true in galaxies as the various global scaling relations (e.g. Tully-Fisher, and even the X-ray behaviour in clusters of galaxies) attest.

The agreement in qualitative behaviour between the scale invariant model and that based on  numerical integration and detailed physical assumptions, is   reassuring. It suggests that  the qualitative behaviour is somewhat insensitive to the detailed physics underlying the model. One sees this also in approximations to the numerical  studies \citep{Chamandy2014b}. However there are some differences. Our time behaviour consists of a power law or exponential growth plus a pattern rotation. There is no predicted intrinsic oscillation as in \citet{Brandenburg1992}, although in our model the projected structure can change relative to a fixed line of sight due to magnetic pattern rotation. This oscillation might be difficult to distinguish from higher order modes. It should be noted that (see Fig.~\ref{fig:rotatingcuts} and Appendix~\ref{app:A}) that our model can produce RM reversals even in axial symmetry due to pitch angle effects. However the self-similarity also restricts the variation of parity with latitude (it happens only once), which may be a distinguishing feature. It is possible that both types of reversals ($m=0$ and $m=n$) occur in combination. Our best fits, in fact, require this.

\section{Conclusions}
\label{section:conclusions}

Remarkable  RM reversals in sign can be seen in the northern halo of RM maps of NGC~4631 as seen in figure \ref{fig:Carolina} and \citet{Carolina2018}.   We solve the classical  dynamo equations under the assumption of scale invariance, and we search for rotating logarithmic spiral modes projected on cones. The three dimensional magnetic fields also have strong poloidal components that appear to loop over the projected spirals near the disk. The model allows for corresponding velocity fields representing accretion onto the disk, outflow from the disk, and rotation-only in a disk pattern frame and we search for solutions for each case.  Our models produce  magnetic fields and consequently RM sign reversals when viewed edge-on. RM maps are created using a Faraday screen and are scaled to amplitude of the observed maps.  Residual images are then made and used to compare how well the different models fit the data.  Solutions for rotation-only cases, in general, did not fit the observations of NGC~4631 well.  Outflow models provided a reasonable fit to the magnetic field structure, but the best results are found using accretion models for the specified region (boxed in Fig.~\ref{fig:Carolina}).

\section*{Acknowledgements}

This work has been supported by a Queen Elizabeth II Scholarship in Science and Technology to AW from the Province of Ontario and Queen's University.  JI wishes to thank the Natural Sciences and Engineering Research Council of Canada for a Discovery Grant.  



\bibliographystyle{mnras}
\bibliography{references} 



\appendix
\section{General Results and Observational Expectations}
\label{app:A}

\begin{table}
\begin{tabular}{ll}
\hline
Parameter  & Physical Interpretation                                                                                                           \\ \hline \hline
$u, v, w$  & Scaled cylindrical velocity components                                                                                    \\
$\epsilon$ & Fixes rate of rotation of magnetic field in time                                                                                  \\
$q$        & Used to define spiral pitch angle. Pitch angle is\\ & found as $\arctan(1/q)$                          \\
$T$        & Time variable                                                                                                                     \\
$m$        & Spiral mode                                                                                                                       \\
$C1, C2$   & Boundary conditions for the magnetic field                                                                                        \\
$a$        & Similarity class, defines globally conserved \\ & quantity (See table ($\ref{table:a})$) \\ \hline
\end{tabular}
\caption{Physical interpretations of parameters used.}
\label{table:parameters}
\end{table}

In this section we display observational expectations from the magnetic fields produced from these dynamos.  We begin by summarizing the different variables found in these solutions and their physical interpretation (see Table~\ref{table:parameters}) and then move on to specific cases.  The images presented in this section are RM maps that are obtained by observing the galaxy as though it were face-on or edge-on. 

The parameter $a$, found in Eqn.~\ref{eq:delta}, is the 'similarity class' of the model.  This parameter represents the dimensions of a globally conserved quantity in the solutions.  This is discussed in greater detail in \citet*{HWI2018} as well as Sect.~\ref{section:spiral}.  A summary of different similarity classes and their possible identifications can be found in Table~\ref{table:a}.   

\begin{table}
\begin{tabular}{lll}
\hline
a   & Dimension of X  & Possible Identification            \\ \hline \hline
0   & $T^{q}$         & Angular velocity if $q=-1$         \\
1   & $L^n/T^n$       & Linear velocity if $n=1$           \\
3/2 & $L^{3n}/T^{2n}$ & Keplerian orbits if $n=1$          \\
2   & $L^{2n}/T^{n}$  & Specific angular momentum if $n=1$ \\
3   & $L^{3n}/T^{n}$  & Magnetic Flux if $n=1$             \\ \hline
\end{tabular}%
\caption{\textbf{Self  Class Identification} \hspace{\textwidth}$^{\textbf{a}}$Recall that magnetic field and velocity have the same dimensions when the field is divided by the square root of an arbitrary density.\hspace{\textwidth} $^{\textbf{b}}$Recall that, generally, $a \equiv \alpha / \delta = p/q$, where the globally conserved quantity, X, has dimensions [X] = $L^{\textbf{p}}/T^{\textbf{q}}$}
\label{table:a}
\end{table}

\begin{figure*}
\begin{tabular}{ccc} 
\rotatebox{0}{\scalebox{0.45} 
{\includegraphics{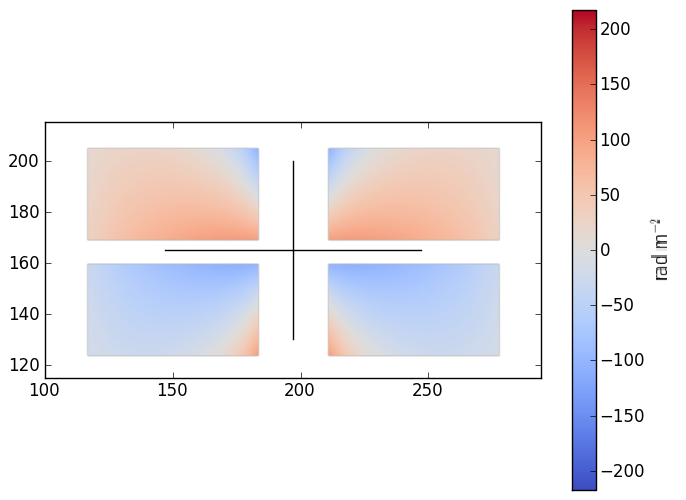}}}&
\rotatebox{0}{\scalebox{0.45} 
{\includegraphics{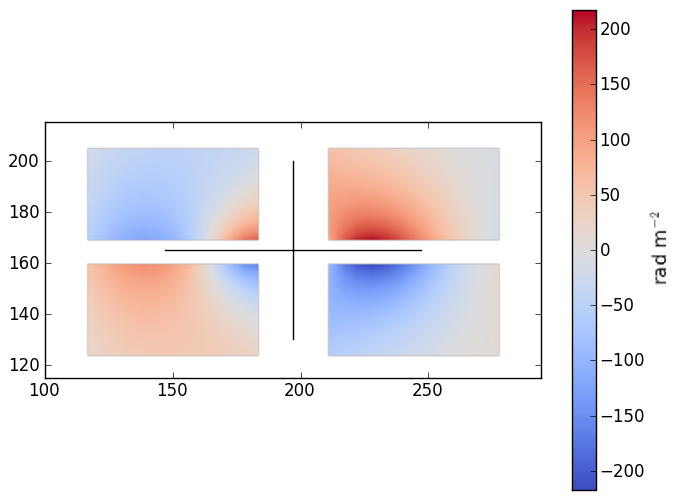}}}&
\rotatebox{0}{\scalebox{0.45} 
{\includegraphics{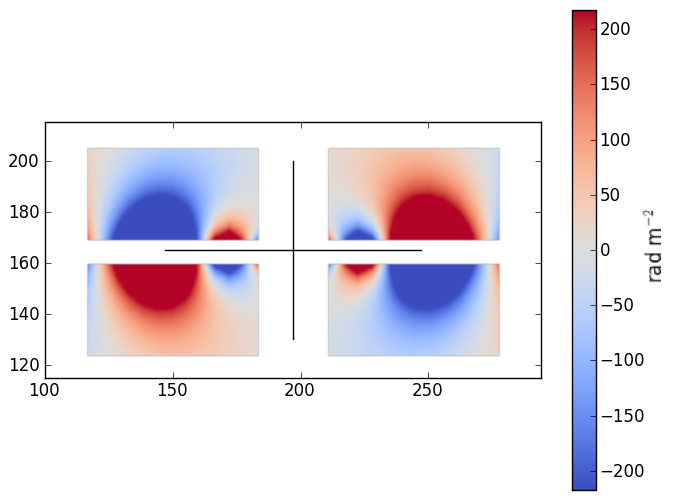}}}\\
\rotatebox{0}{\scalebox{0.33} 
{\includegraphics{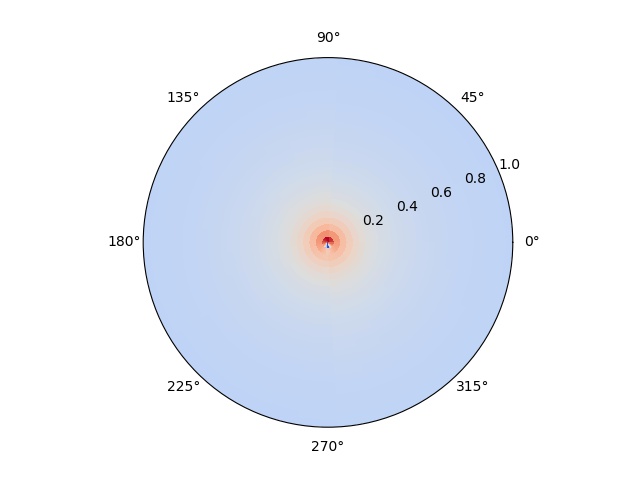}}}&
\rotatebox{0}{\scalebox{0.33} 
{\includegraphics{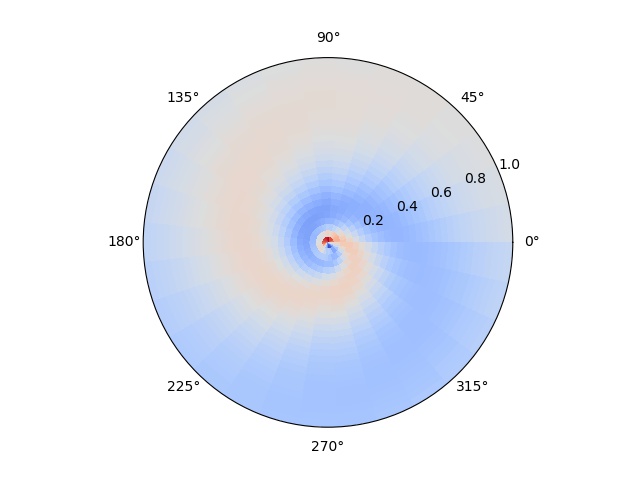}}}&
\rotatebox{0}{\scalebox{0.33} 
{\includegraphics{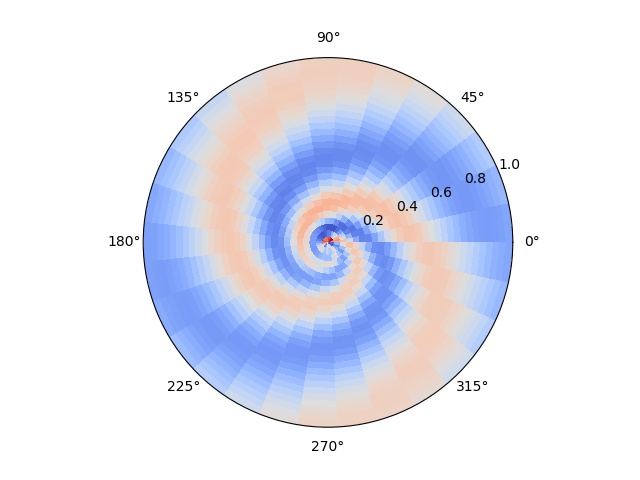}}}\\
\end{tabular}
\caption{We show edge-on (top row) and face-on (bottom row) RM screens for the parameter vector $\{m$, $q$, $\epsilon$, $u$, $v$, $w\}$ = $\{m, 2.5, 0.0, 0.0, 0.0, 2.0\}$.  This is an example of outflow-only from the rotation frame.  Parameter $m$ is allowed to vary from left to right.  In the leftmost column $m=0$, in the middle column $m=1$, and in the rightmost column $m=2$.  The radius in units of the galactic radii is shown on the face-on figures.  The number of magnetic spirals can be seen increase in the face-on case with the number of arms corresponding to the value of $m$.  This arms can be seen as reversals in the edge-on screens.}
\label{fig:m}
\end{figure*}

The parameter $m$ is used in these spiral solutions to indicate the spiral mode, that is the number of spirals appearing in the solution.  In Eqn.~\ref{eq:fieldform} solutions for the magnetic field potential ${\bf \bar A}$ are searched for in the complex form ${\bf \bar A}(R,\Phi,Z)={\bf \tilde A}(\zeta)e^{im\kappa}$.  Fig.~\ref{fig:m} shows edge-on (top row) and face-on (bottom row) RM screens produced for different values of m when other parameters are kept constant. A number of projected magnetic spirals corresponding to the value of m can clearly be seen.  The number of sign reversals in the edge-on case increases with increasing m however it should be noted that counting the number of reversals alone cannot determine the value of m seen.  The spiral pitch angle discussed later can also cause the projected spiral structure to wrap more tightly or loosely causing more or less reversals to be seen in the edge-on case.  

Solutions of the dynamo equations for different values of m are independent of one another, however this does not preclude that multiple solutions may be present.  Solutions with the same parameter vector apart from various values of m can be combined together (e.g. solutions for $m=0$, $m=1$, and $m=2$ can be combined to produce a new RM map). 

\begin{figure*}
\begin{tabular}{ccc} 
\rotatebox{0}{\scalebox{0.45} 
{\includegraphics{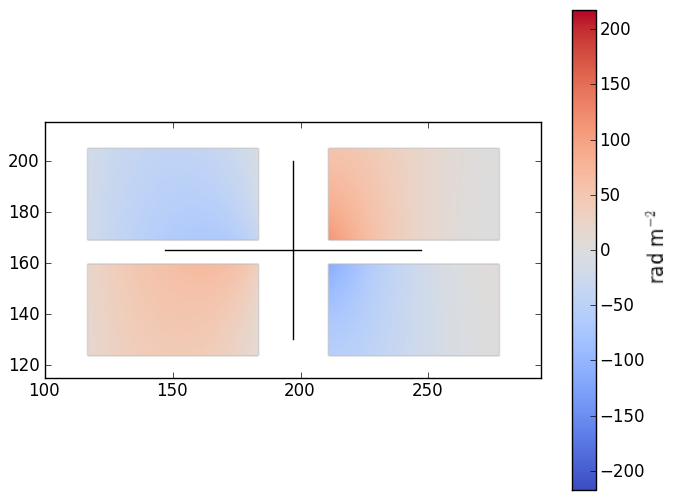}}}&
\rotatebox{0}{\scalebox{0.45} 
{\includegraphics{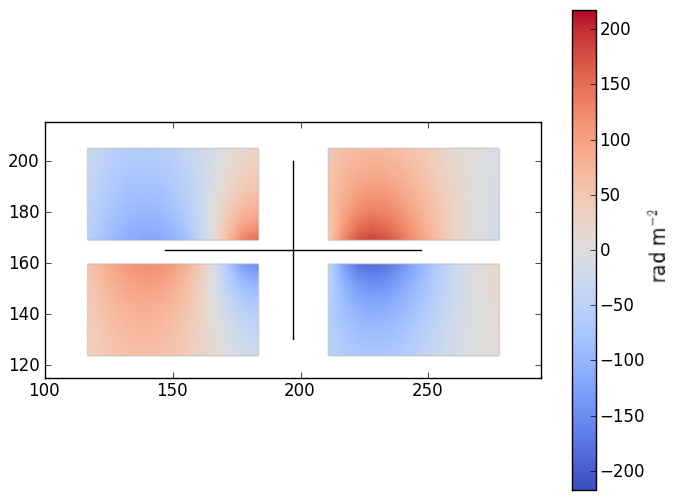}}}&
\rotatebox{0}{\scalebox{0.45} 
{\includegraphics{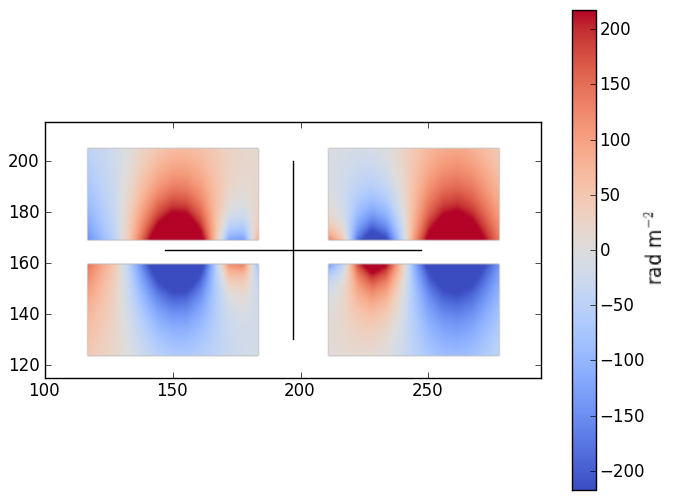}}}\\
\rotatebox{0}{\scalebox{0.33} 
{\includegraphics{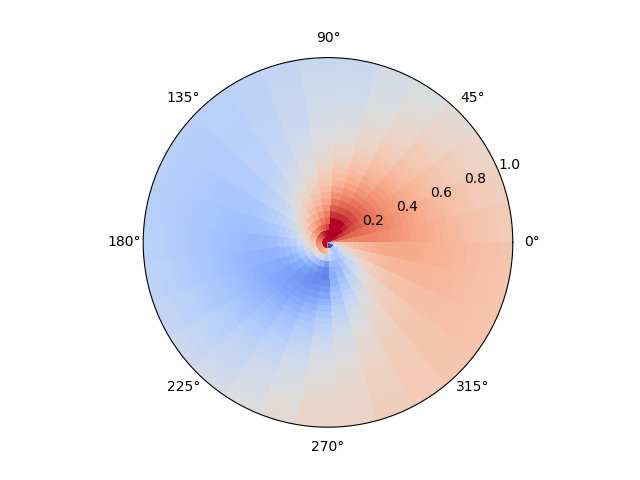}}}&
\rotatebox{0}{\scalebox{0.33} 
{\includegraphics{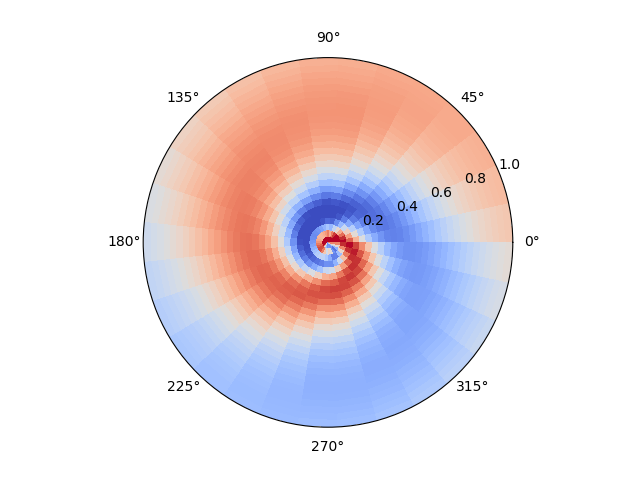}}}&
\rotatebox{0}{\scalebox{0.33} 
{\includegraphics{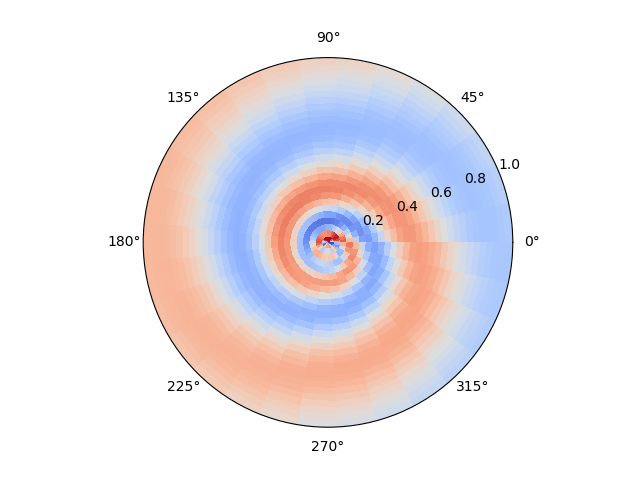}}}\\
\end{tabular}
\caption{We show edge-on (top row) and face-on (bottom row) RM screens for the parameter vector $\{m$, $q$, $\epsilon$, $u$, $v$, $w\}$ = $\{1, q, 0.0, 0.0, 0.0, 4.0\}$.  The radius in units of the galactic radii is shown on the face-on figures.  This is an example of outflow-only from the rotation frame.  Parameter $q=1.0,2.5,4.9$ from left to right respectively.  The number of spirals remains constant however becomes more tightly would as $q$ increase.  This results in increasing the number of reversals seen in the edge-on case.  Scaling depends on an arbitary multiplicative constant.}
\label{fig:q}
\end{figure*}

Parameters $\epsilon$ and $q$ appear in Eqn.~\ref{eq:kappa} where they are used to define rotating logarithmic spiral forms.  Parameter $q$ represents the pitch angle of the spiral solution.  The pitch angle can be found as $\arctan(1/q)$.  Fig.~\ref{fig:q} shows face-on and edge-on rotation measure screens for the same parameter vector with varying $q$.  As can be seen a higher $q$ decreases the angle of the magnetic spirals.  In the edge-on case a lower $q$ (higher pitch angle) causes the spirals to become more tightly wound and produces more reversals across the galaxy halo.  The number of reversals seen in the edge-on case depends on both the spiral mode as well as the pitch angle in these solutions.

\begin{figure*}
\begin{tabular}{ccc} 
\rotatebox{0}{\scalebox{0.33} 
{\includegraphics{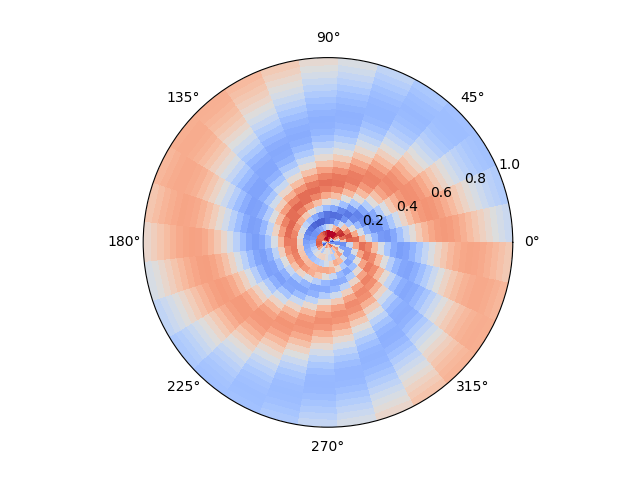}}}&
\rotatebox{0}{\scalebox{0.33} 
{\includegraphics{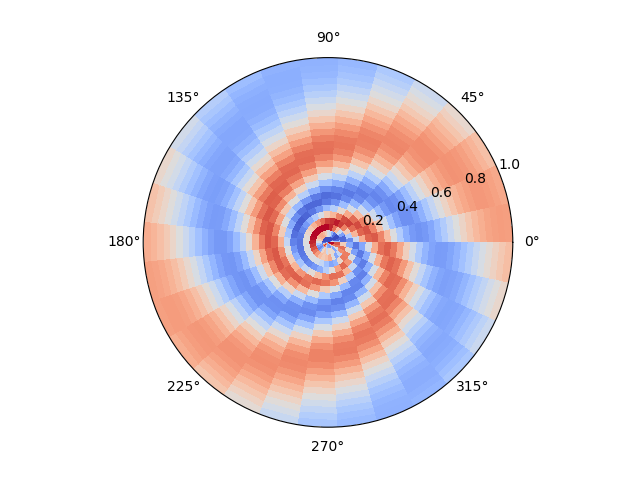}}}&
\rotatebox{0}{\scalebox{0.33} 
{\includegraphics{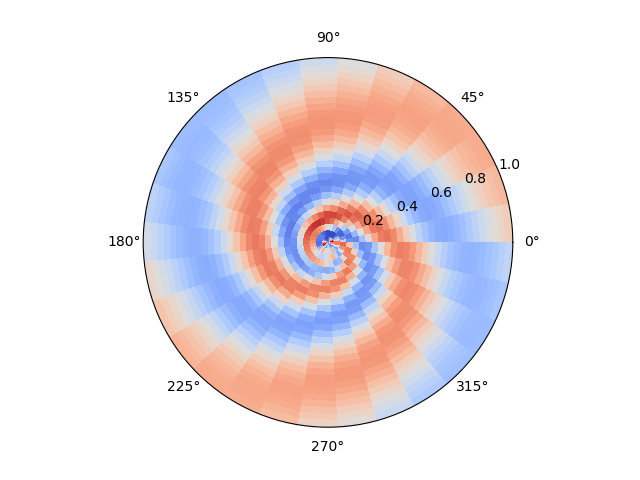}}}\\
\end{tabular}
\caption{We show face-on RM screens for the parameter vector $\{m$, $q$, $\epsilon$, $u$, $v$, $w\}$ = $\{2, 2.5, \epsilon, 0.0, 0.0, 4\}$ with $\epsilon=-0.5,0.0,0.5$ from left to right respectively.  The radius in units of the galactic radii is shown on the face-on figures and scaling depends on an arbitary multiplicative constant.  This is an example of an outflow model.  The spiral pattern can be seen rotating as $\epsilon$ is varied.  The parameter $\epsilon$ can be used to simulate rotation with time of the spiral pattern.}
\label{fig:epsilon}
\end{figure*}

The parameter $\epsilon$ is a number that  fixes  the  rate  of  rotation  of  the  magnetic  field  in  time.  By varying $\epsilon$ one can rotate the field emulating its rotation with time.  This is seen in Fig.~\ref{fig:epsilon} where magnetic structure can be seen rotating as epsilon is increased.  

Parameters $u$, $v$, $w$ are scaled cylindrical velocity components where $u$ is in the $r$ direction, $v$ is in the $\theta$ direction, $w$ is in the z direction.  These are discussed further in the next section.

\subsection{Outflow or Accretion in the Pattern Reference Frame}\label{section:outflowaccretion}

As explained in Sect.~\ref{section:bi-symmetric}, we will restrict ourselves to solutions where $a=1$ and $u=v=0$ to study outflow from, and accretion onto, the galactic disk.  For these solutions $w$ is allowed to vary and represents the relative amount of inflow/outflow onto the disk.  A positive $w$ indicates outflow and a negative $w$ indicates accretion.  

\begin{figure*}
\begin{tabular}{ccc} 
\rotatebox{0}{\scalebox{0.45} 
{\includegraphics{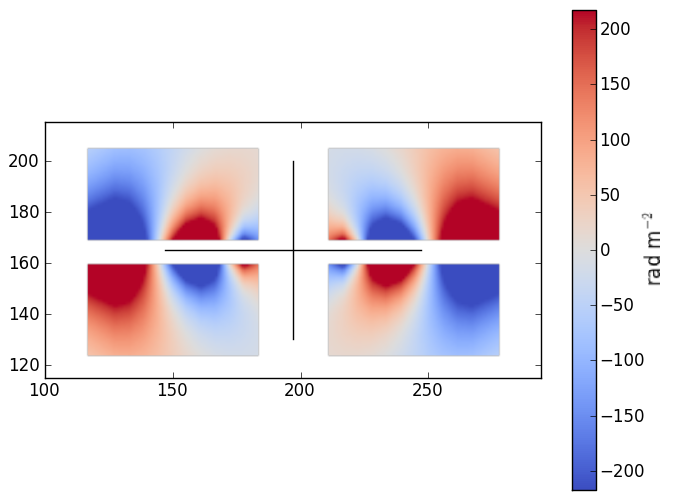}}}&
\rotatebox{0}{\scalebox{0.45} 
{\includegraphics{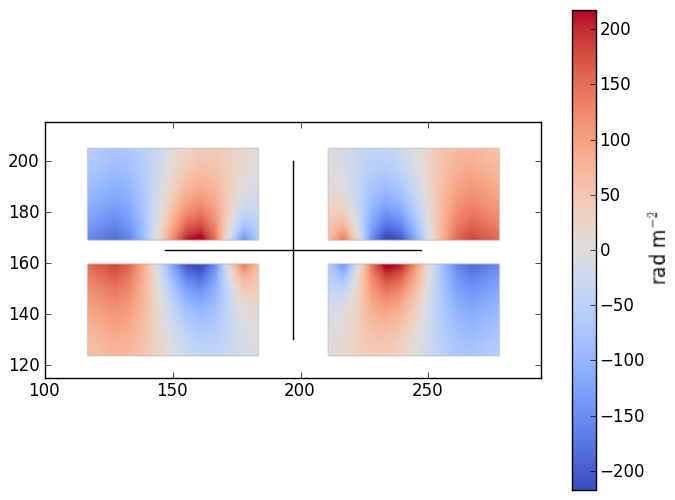}}}&
\rotatebox{0}{\scalebox{0.45} 
{\includegraphics{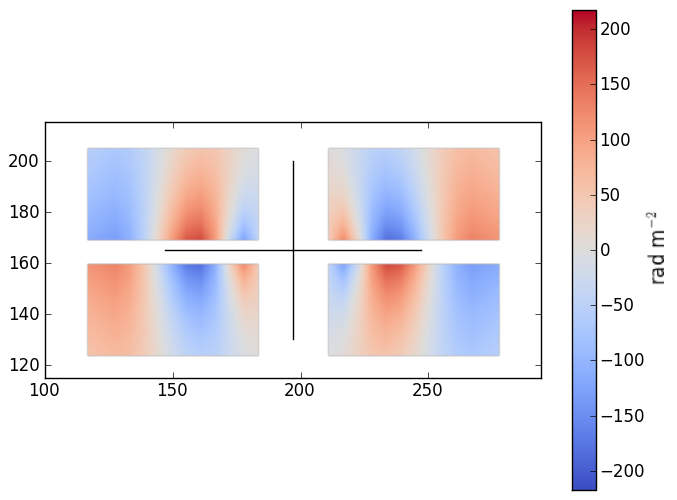}}}\\
\end{tabular}
\caption{We show edge-on RM screens for the parameter vector $\{m$, $q$, $\epsilon$, $u$, $v$, $w\}$ = $\{2, 2.5, 0.0, 0.0, 0.0, w\}$ with $w=-2,-4,-5$ from left to right respectively.  Scaling depends on an arbitary multiplicative constant.  This is an example of an accretion model.  All maps show the same structure however the reversals become more spread out in the vertical direction as the wind speed increases.  Reversal patterns also become more straight and less curved with increasing wind speed.}
\label{fig:w}
\end{figure*}

In Fig.~\ref{fig:w} $w$ is varied for an accretion case where all other parameters are kept constant.  As can be seen in this figure the strength of the reversals decreases as the wind speed increases, with a stronger wind producing more well defined reversals.  These reversals also have a more vertical structure with less curvature to the shape of the reversals.  In the $w=-2$ case the reversals have a more curved structure, displaying a more kidney bean like structure, while in the $w=-5$ case the reversals display a much more vertical structure.  

\begin{figure*}
\begin{tabular}{ccc} 
\rotatebox{0}{\scalebox{0.45} 
{\includegraphics{m1.png}}}&
\rotatebox{0}{\scalebox{0.45} 
{\includegraphics{m2.png}}}&
\rotatebox{0}{\scalebox{0.45} 
{\includegraphics{m3.png}}}\\
\rotatebox{0}{\scalebox{0.45} 
{\includegraphics{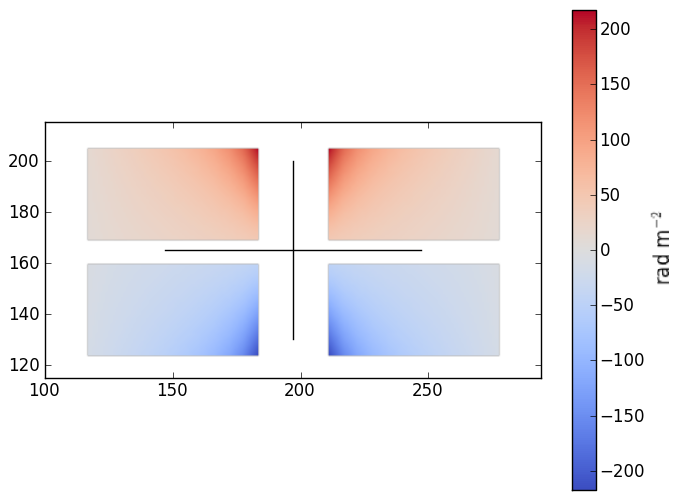}}}&
\rotatebox{0}{\scalebox{0.45} 
{\includegraphics{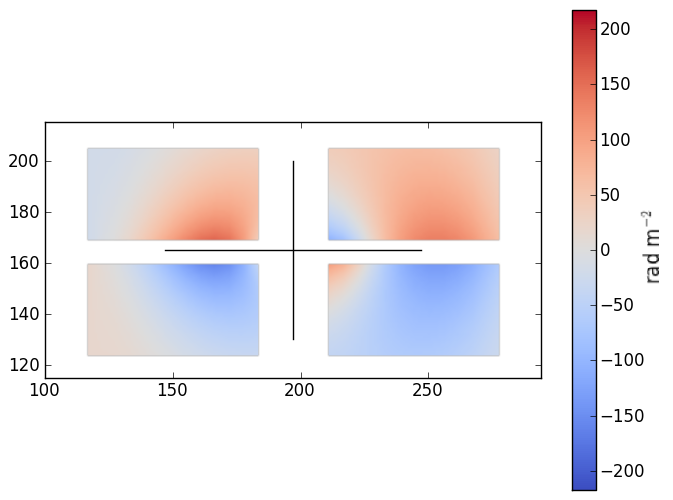}}}&
\rotatebox{0}{\scalebox{0.45} 
{\includegraphics{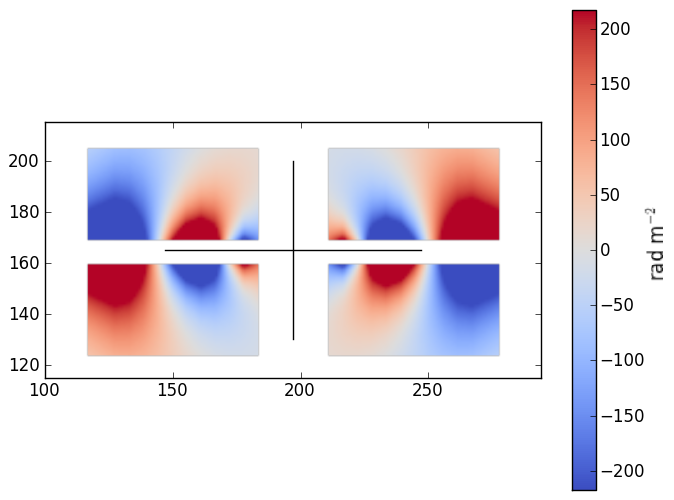}}}\\
\end{tabular}
\caption{We show edge-on RM screens (with arbitary scaling) for the parameter vector $\{m$, $q$, $\epsilon$, $u$, $v$, $w\}$ = $\{m, 2.5, 0.0, 0.0, 0.0, \pm2.0\}$ where $w=+2$ for the top row and $w=-2$ for the bottom row.  The parameter is is varied as $m=0,1,2$ from left to right respectively.  This figure therefore shows the same solution for outflow in the top row and inflow in the bottom row.  Solutions are in general similar however not necessarily the same.}
\label{fig:inout}
\end{figure*}

Solutions for inflow (accretion) and outflow (winds) in general display similar RM maps and can be difficult to distinguish.  Fig.~\ref{fig:inout} shows different cases of inflow and outflow solutions for edge-on cases.  All solutions display similar spiral reversals in $m \neq 0$ cases seen as reversals across the halo in edge-on galaxies.  Outflow versus inflow solutions with the same parameter sets are in general very similar, they however may not display precisely the same patterns.  For example in Fig.~\ref{fig:inout} the images in the top row are for the same parameter set as the images in the bottom row with $m=0,1,2$ from left to right respectively except the velocity in the $w$ direction is opposite in sign.  While the outflow solution for $m=0$ (top left image) shows a field reversal, the inflow solution for $m=0$ (bottom row) does not.  

\begin{figure*}
\begin{tabular}{ccc} 
\rotatebox{0}{\scalebox{0.45} 
{\includegraphics{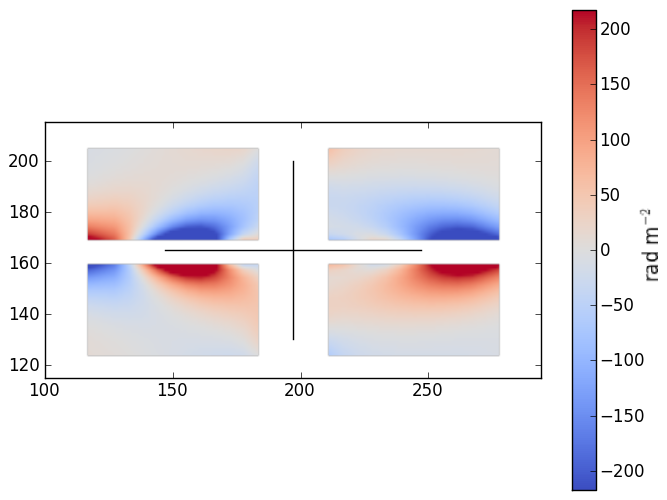}}}&
\rotatebox{0}{\scalebox{0.45} 
{\includegraphics{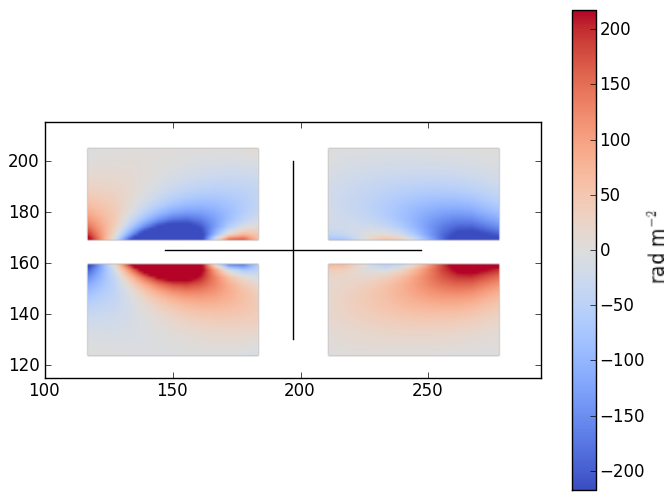}}}&
\rotatebox{0}{\scalebox{0.45} 
{\includegraphics{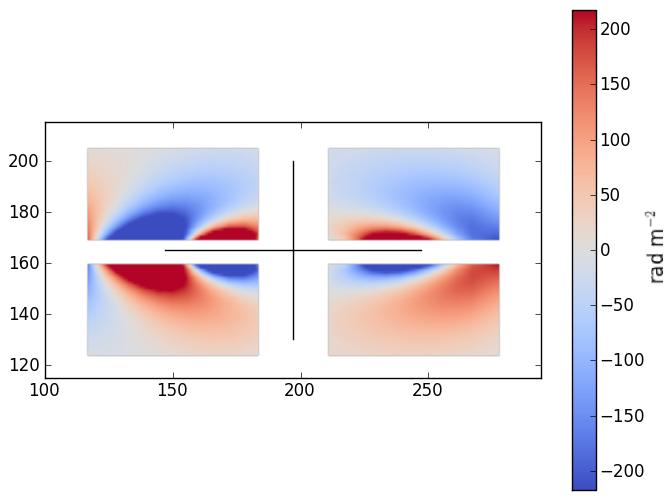}}}\\
\end{tabular}
\caption{We show edge-on RM screens for the parameter vector $\{a$, $m$, $q$, $\epsilon$, $u$, $v$, $w\}$ = $\{a, 2, 2.5, -1.0, 0.0, 1.0, 0.0\}$.  This is an example of rotation-only in the pattern frame.  Parameter $a$ is allowed to vary from left to right.  In the leftmost column $a=0$, in the middle column $a=1$, and in the rightmost column $a=2$.  No clear pattern can be consistently discerned from variations of $a$.  Images with the same parameter vector with $a$ varied are visually similar to one another and do not change drastically.  Note the scaling depends on an arbitrary multiplicative constant.}
\label{fig:a}
\end{figure*}

\subsection{Rotation-Only in the Pattern Reference Frame}

In this subsection we restrict ourselves to solutions where there is rotation-only in the pattern frame by  setting $u=w=0, v \neq 0$.  Unlike the previous subsection this allows $a$ to be arbitrary and a parameter of the solutions.  In Fig.~\ref{fig:a} $a$ is varied while all other parameters are kept constant.  No discernible pattern can be distinguished between varying $a$ as a parameter and the solutions appear to be independent from one another.

Solutions appear to contain strong kidney bean shaped reversals near the disk and reversals are not seen to be linear in height above the disk, rather they curve towards the center.  These solutions are distinguishable from the inflow/outflow case by these strong kidney bean shaped reversals as well as solutions being closer to the disk.  Reversals in the rotation-only case appear to be bigger in radii than in the inflow/outflow case.  Outside of the strong reversal regions little Faraday rotation in usually seen.


\bsp	
\label{lastpage}
\end{document}